\documentclass[preprint]{aastex}

\begin{document}

\makeatletter
\newcommand{\rmnum}[1]{\romannumeral #1}
\newcommand{\Rmnum}[1]{\expandafter\@slowromancap\romannumeral #1@}
\makeatother

\title{Spitzer observations of supernova remnants: \Rmnum{2}. Physical conditions and comparison with HH7 and HH54}

\author{Yuan Yuan and David A. Neufeld}

\affil{Department of Physics and Astronomy,
Johns Hopkins University, 3400 North Charles Street, Baltimore, Maryland 21218}

\begin{abstract}

We have studied the shock-excited molecular regions associated with four
supernova remnants (SNRs) | IC443C, W28, W44 and 3C391 | and two Herbig-Haro
objects, HH7 and HH54, using $\it{Spitzer}$'s Infrared Spectrograph (IRS).
The physical conditions within the
observed areas (roughly
$\sim~1^{'}\times~1^{'}$ in size) are inferred from spectroscopic data
obtained from IRS and from the Short (SWS) and Long (LWS) Wavelength
Spectrometers onboard the {\it Infrared Space Observatory} ({\it
ISO}), together with photometric data from $\it{Spitzer}$'s
Infrared Array Camera (IRAC).

Adopting a power-law distribution for the gas temperature in the
observed region, with the mass of gas at temperature $T$ to $T+dT$
assumed proportional to $T^{-b} dT$, the H$_{2}$ S(0) to S(7)
spectral line maps obtained with IRS were used to constrain the gas
density, yielding estimated densities $n$(H$_2$) in the range of
$\sim$ 2 -- 4$\times10^{3}$ cm$^{-3}$. The excitation of H$_2$ S(9)
to S(12) and high-$J$ CO pure rotational lines, however, require
environments several times denser. The inconsistency among the
best-fit densities estimated from different species can be explained
by density fluctuations within the observed regions. The best-fit
power-law index $b$ is smaller than the value 3.8 predicted for a
paraboloidal C-type bow shock, suggesting that the shock front has a
``flatter'' shape than that of a paraboloid. The best-fit parameters
for SNRs and Herbig-Haro objects do not differ significantly between
the two classes of sources, except that for the SNRs the
ortho-to-para ratio (OPR) of hot gas ($T>$ 1000 K) is close to the
LTE value 3, while for HH7 and HH54 even the hottest gas exhibits an
OPR smaller than 3; we interpret this difference as resulting from
environmental differences between these classes of source, molecular
material near SNRs being subject to stronger photodissociation that
results in faster para-to-ortho conversion. Finally, we mapped the
physical parameters within the regions observed with IRS and found
that the mid-lying H$_2$ emissions | S(3) to S(5) | tend to trace
the hot component of the gas, while the intensities of S(6) and S(7)
are more sensitive to the density of the gas compared to S(3) to
S(5).

\end{abstract}


\section{Introduction}

Interstellar shocks generated by violent stellar activities, such as
supernova explosions and protostellar outflows, have profound
effects on the surrounding interstellar medium. Shocks propagating into
dense molecular clouds can heat the gas to several hundred or
several thousand Kelvin and produce rich spectra in the infrared
spectral region. Fast shocks with speeds larger than 40 km s$^{-1}$
are usually dissociative; they destroy molecules and ionize atoms,
generating strong atomic fine-structure emissions. On the other
hand, most molecules survive in slow shocks. Heating excites various
molecular species via collisional processes, causing the gas to
glow. A large number of molecular line features have been observed
in association with shock-affected areas, including cooling lines
from H${_2}$, CO, HD and H${_2}$O. Due to its ability to reveal
species that are difficult to observe within cold quiescent gas, a
shock wave serves as a ``searchlight'' for the physical structure of
molecular clouds.

Moreover, shocks alter the chemical composition of gas by driving
many endothermic reactions, one of which is the conversion of para
molecular hydrogen to ortho hydrogen.  In previous studies, it has
been found that many sources exhibit H${_2}$ ortho-to-para ratio
(hereafter OPR) markedly less than the equilibrium value $\sim 3$
(Neufeld et al. 1998; Cabrit, et al. 1999; Neufeld et al. 2006,
hereafter N06; Neufeld et al. 2007, hereafter N07). Adopting a two
component model containing a mixture of warm and hot gas, N06 and
N07 found that for the sources we are studying in this paper --
IC443C, W28, W44, 3C391, HH7 and HH54 -- the OPR values in the warm
gas component ($T \sim 300-600$ K) are 2.42, 0.93, 1.58, 0.65, 0.21
and 0.41--0.48 respectively. They proposed that the non-LTE OPR
values obtained may correspond to the temperature at an earlier
epoch, owing to the low efficiency of para-to-ortho conversion in
non-dissociative shocks; this then can provide us useful information
on the evolution timescale.

In this paper we investigate molecular shocks associated with four
supernova remnants -- IC443C, W28, W44 and 3C391 and two Herbig-Haro
objects -- HH7 and HH54. All six sources have extensive evidence for
interaction with molecular clouds provided by multi-wavelength
observations. A brief description of these sources is given below.

The four bright supernova remnant sources IC443, W28, W44 and 3C391
have provided excellent laboratories for the study of the
interaction between SNR shocks and surrounding molecular gas.  W28,
W44 and 3C391 are prototypes of the ``mixed-morphology'' class,
whose centrally concentrated X-ray morphologies contrast with the
shell-like radio emission (Rho et al. 1994; Rho \& Peter 1996; Rho
\& Peter 1998). The radio and X-ray morphology of IC443 also shows
similarities to the mixed-morphology class, although with additional
X-ray components. X-ray observations show the four remnants to be
filled with a large amount of hot gas, with density $n~\sim$ 1 -- 10
cm$^{-3}$ according to the radiative model of Harrus et al. (1997)
and Chevalier (1999); this implicates an interaction with relatively
dense environments | probably intercloud gas. More convincing
evidence for the interaction with dense clouds arises directly from
the detection of various excited molecules at longer wavelengths. As
summarized by Reach et al. (2005) and N07 for each individual source
in detail, this evidence includes broad line emissions in the
millimeter-wave region; OH maser emission; near-infrared H$_2$
emission observed mainly by ground-based observatories; and
mid-infrared emissions from various molecular species including
H$_{2}$, CO, HD, H$_{2}$O, PAH, as well as atomic fine structure
lines. Most of the mid-infrared spectra have been recently provided
by the {\it Infrared Space Observatory} ({\it ISO}) and the {\it
Spitzer Space Telescope}. All the evidence mentioned above has been
obtained for all four SNRs.

IC443C is one of the four areas in IC443 marked by DeNoyer (1978),
where bright condensations of HI have been seen. Near infrared
observations reveal IC443 is evolving in a complex environment, with
the northeast rim dominated by atomic fine structure lines including
[FeII], [OI], etc. , and the southern ridge dominated by molecular
line emissions, especially H$_{2}$ emissions (Rho et al. 2001).
Rho et al. (2001) argued for the existence of fast shocks ($V_s$ $\sim$
100 km s$^{-1}$) propagating in moderately dense gas ($n \sim$ 10  --
10$^{3}$~ cm$^{-3}$) within the northestern rim, and slow shocks
($V_s$ $\sim$ 30 km s$^{-1}$) in the denser ($n \sim$ 10$^{4}$~
cm$^{-3}$) southern part. IC443C coincides with the peak of the H$_{2}$
rovibrational emission in the southern rim.

 HH7 is one of a chain of Herbig-Haro objects located in the
star-forming region NGC 1333, separated by HH8 -- 11 from the
probable protostar SVS 13 (Strom, Vbra \& Strom, 1979). It has been
investigated extensively through optical and infrared observations
as summarized by N06. HH7 exhibits a well defined bow shock in the
infrared. Smith et al.\ (2003) studied the image of H$_2$
rovibrational emissions and found it can be well modeled by a
paraboloidal bow shock with speed $\sim$ 55 km s$^{-1}$ and preshock
density $\sim$ 8$\times$10$^{3}$ cm$^{-3}$. HH54 is located at the
edge of a star forming cloud Cha II (Hughes \& Hartigan 1992) and
consists of a complex of arcsecond scale bright knots (Graham \&
Hartigan 1988). The presence of warm molecular hydrogen was first
reported by Sandell et al. (1987) and confirmed by Gredel (1994)
with near-infrared observations. More infrared data come from
observations with $\it{ISO}$ and $\it{Spitzer}$ (Cabrit et al. 1999;
Neufeld et al. 1998; N06; Giannini et al. 2006) with the detection
of several molecular features from H$_{2}$, HD, CO, H$_{2}$O, as
well as many ionic lines.

The structure of this paper is as follows: the observational data we
employed are discussed in Section 2; our analysis method is
described in Section 3 and Appendix A along with the shock model;
results for each individual source and a discussion are presented in
Section 4 and 5; Section 6 serves as a brief summary of the paper.

\section{Observations}

In this paper we analyze the physical conditions in interstellar
areas affected by interstellar shocks with the use of spectroscopic data obtained from the
Infrared Spectrograph (IRS) on board {\it Spitzer} and two
spectrometers on board {\it ISO}, as well as photometric data from
the {\it Spitzer}'s Infrared Array Camera (IRAC).

\subsection{IRS observations of H$_{2}$ and HD}

Spanning a wide wavelength range from 5.2 to 37 microns, IRS on {\it
Spitzer} provides access to pure rotational H$_{2}$ lines from $v$ =
0 -- 0 S(0) to S(7) and a variety of fine structure lines including
[Fe II], [S I], [Ne II] etc.  Spectral maps of the six sources were
obtained using the Short-Low (SL), Short-High (SH) and Long-High
(LH) modules of IRS. Most of the data we employ here come from
observations performed as part of the Cycle 2 General Observer
Program, in which the IRS slit was stepped one-half of its width
perpendicular to its long axis and 4/5 of its length parallel to the
axis to map fields of size $\sim~1^{'}\times~1^{'}$.  In Cycle 3, we
obtained additional LH data with integration times a factor of 14 --
60 longer than those obtained in Cycle 2 for the H$_2$ S(0), HD R(3)
and R(4) lines toward IC443C, HH7 and HH54, providing an improved
signal-to-noise ratio. These Cycle 3 LH observations were carried
out on 2007 April 22, 2007 November 11, 2007 May 1 and 3, and 2007
October 1 respectively for IC443C, HH54, and HH7. For the new
observations $1^{'}\times~1^{'}$ regions were mapped by stepping IRS
slit one-half its width perpendicular and 1/5 of its length parallel
to its long axis.  A detailed discussion of the data reduction
procedures we adopted is given in N06 and N07.

In addition to H$_2$, the two HD rotational lines R(3) and R(4)
detected in the IRS LH module toward IC443C, HH54 and HH7
provide us with an extra diagnostic of gas densities and the HD
abundance (N06). A detailed discussion of our HD detections and
abundance measurements will be presented in a future paper, which is
in preparation.  For each source, the line intensities for each H$_2$ and HD
transition are
listed in Table 1, averaged over the rectangular regions enclosed within
the solid line boxes in Figure 1 -- 6 to avoid pixels with poor
signal-to-noise ratios.

\subsection{IRAC observations and comparison with IRS}

In addition to the IRS spectroscopic observations, photometric
observations with IRAC on {\it Spitzer} may also help us in studying
H$_{2}$. With four filters centered at 3.6, 4.5, 5.8 and 8 $\mu$m,
IRAC covers a variety of rovibrational and pure rotational
transitions of H$_{2}$ (Reach et al. 2006; NY08), providing access
to higher excitation levels than those observable by IRS.  More
specifically, IRAC band 1 (covering $\sim$ 3.2 -- 4.0$\mu$m) is
sensitive to the $v$ = 1 -- 0 O(5) to O(7) and $v$ = 0 -- 0 S(13) -
S(17) transitions of H$_2$; band 2 (3.8 -- 5.1 $\mu$m) is sensitive
to the $v$ = 0 -- 0 S(9) to S(12) transitions; band 3 (4.9 to 6.5
$\mu$m) is sensitive to the $v$ = 0 -- 0 S(6) to S(8) transitions;
while band 4 (6.3 -- 9.6 $\mu$m) is sensitive to the $v$ = 0 -- 0
S(4) and S(5) transitions. The contributions of H$_{2}$ line
emissions to each IRAC band are listed in Table 1 of NY08.

\begin{figure}
\includegraphics[scale=0.8]{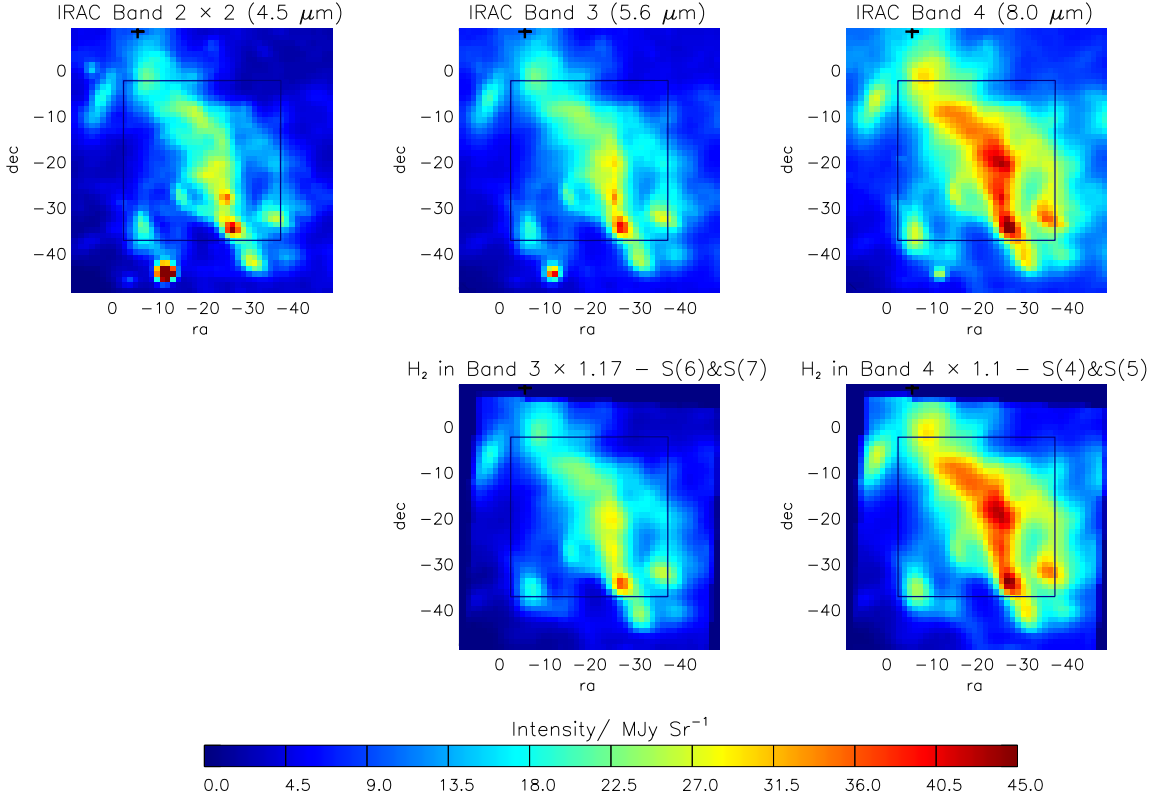}
\caption{The upper panels show the three IRAC bands for IC443C at
4.5, 5.6, 8 $\mu$m and the lower panels show the IRS H$_2$ line
contributions in 5.6 $\mu$m and 8$\mu$m bands calculated using
equations (1) and (2) in NY08. The horizontal and vertical axes
represent offsets in arcsec relative to $\alpha =
6^{h}17^{m}44^{s}.2$ , $\delta= 22^{\circ}21'49''$ (J2000). IRAC
band 3 (5.6 $\mu$m) is attributed mainly to H$_2$ S(6) and S(7)
while IRAC band 4 (8 $\mu$m) is dominated by S(4) and S(5). We
multiply the lower H$_2$ maps by a factor of 1.17 and 1.1
respectively to correct the ``relative uncertainties" exiting
between IRS and IRAC maps. The IRS H$_2$ and HD line intensities in
table 1 are averaged over the region confined by the the solid line
box. The cross in the northeast marks the center of the 80$''$ LWS
beam observing the CO high-lying rotational
lines.}\label{fig:irs&iracIC443}
\end{figure}

After comparing the IRS H$_{2}$ spectral line maps pixel-by-pixel
for IC443C with a IRAC map covering the same region obtained in
Program 68, NY08 found that for this SNR source the IRAC band 3 (5.8
$\mu$m) and band 4 (8 $\mu$m) intensities are contributed almost
entirely by H$_{2}$ $v$ = 0  --  0 S(4) to S(7) line emissions. The
IRS spectrum for IC443C has been presented by N07, averaged over a
Gaussian beam with HPBW $\sim$ 25" centered at $\alpha =
6^{h}17^{m}44^{s}.2$ , $\delta= 22^{\circ}21'49''$ (J2000); this
spectrum indicates that the H$_2$ pure rotational lines are the only
detectable features within the two bandpasses. Moreover, according
to the excitation model of NY08, H$_2$ S(8) emission accounts for
only $\sim$3\% of the observed band 3 intensity. Following the
approach adopted in NY08, we present in Figure
\ref{fig:irs&iracIC443} a comparison between the IRAC band maps and
the spatial distributions of corresponding band intensities
contributed by IRS-observed H$_{2}$ lines only. These band
intensities were calculated with the use of the IRAC spectral
response functions presented by Fazio et al.\ (2004),  in accord
with equations (1) and (2) in NY08.  We found that the IRS-derived
5.8 $\mu$m and 8 $\mu$m H$_{2}$ maps (based upon the H$_2$ line
strengths) could be brought into excellent agreement with the
observed IRAC band maps if multiplied by correction factors of 1.17
and 1.10 respectively.  A more detailed comparison of the band
intensities at each pixel is given in NY08's Figure 4. The
difference between the IRAC maps and maps derived from IRS H$_{2}$
emissions may be caused by different background measurements and
systematic errors in flux calibration. For IC443C, most IRS-mapped
regions are free of pollution from point sources.

According to the excitation model, the IRAC 4.5 $\mu$m band should
also be attributed mainly to H$_{2}$ $v$ = 0 -- 0 S(9) to S(12) line
emissions. Significant contributions from other possible sources
have been ruled out by ground-based spectroscopic observations for
IC443C, including atomic hydrogen recombination lines (Br$\alpha$ at
4.05$\mu$m and Pf$\beta$ at 4.65 $\mu$m) (Burton et al. 1988) , [Fe
II] fine structure lines and CO $v$ = 1 -- 0 rovibrational
transitions (Richter et al. 1995). Similarly, for IC443C, we expect
the 3.6 $\mu$m channel to be dominated by H$_{2}$ $v$ = 1 -- 0
rovibrational transitions (mainly $v$ = 1 -- 0 O(5)). PAH continuum
emissions, which are not detectable towards IC443C within the longer
wavelength range in IRS, are expected to be unimportant in band 1.
On the basis of the analysis above, IRAC band maps of IC443C provide
another probe of the excitation conditions for H$_{2}$. Analysis of
the IRS and IRAC maps for the other five sources will be given
below. For cases where H$_{2}$ emission dominates, the IRAC band 2
brightness is used in conjunction with the spectroscopic data to
constrain the excitation conditions for molecular hydrogen.

\begin{figure}
\includegraphics[scale=0.8]{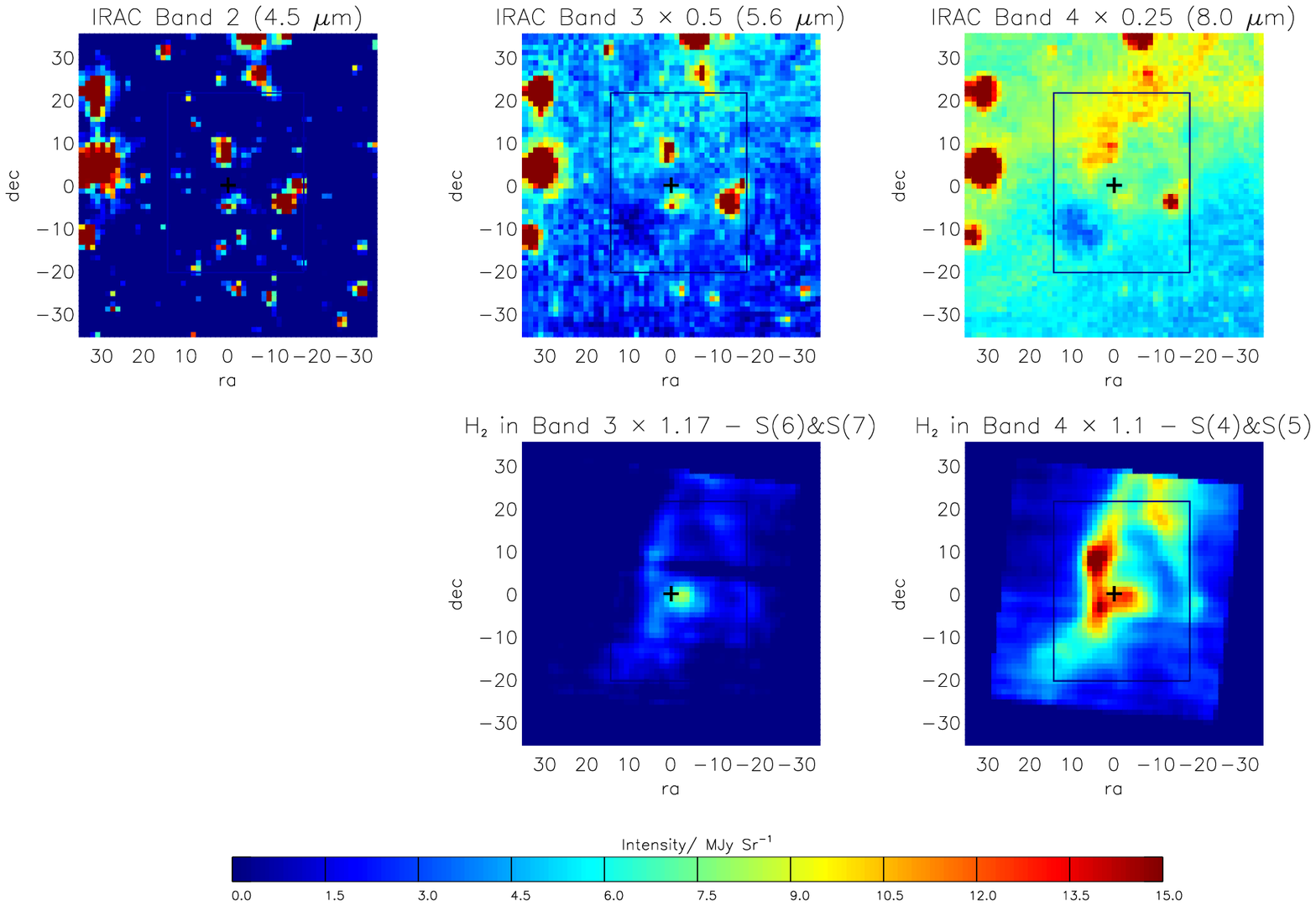}
\caption{Similar as figure 1 but for W28. The (0, 0) position is now
at $\alpha = 18^{h}01^{m}52^{s}.3$ , $\delta= -23^{\circ}19'25''$
(J2000). The center of the $\it{ISO}$ observation (both SWS and LWS)
coincides with the (0,0) position in the
maps.}\label{fig:irs&iracW28}
\end{figure}

\begin{figure}
\includegraphics[scale=0.8]{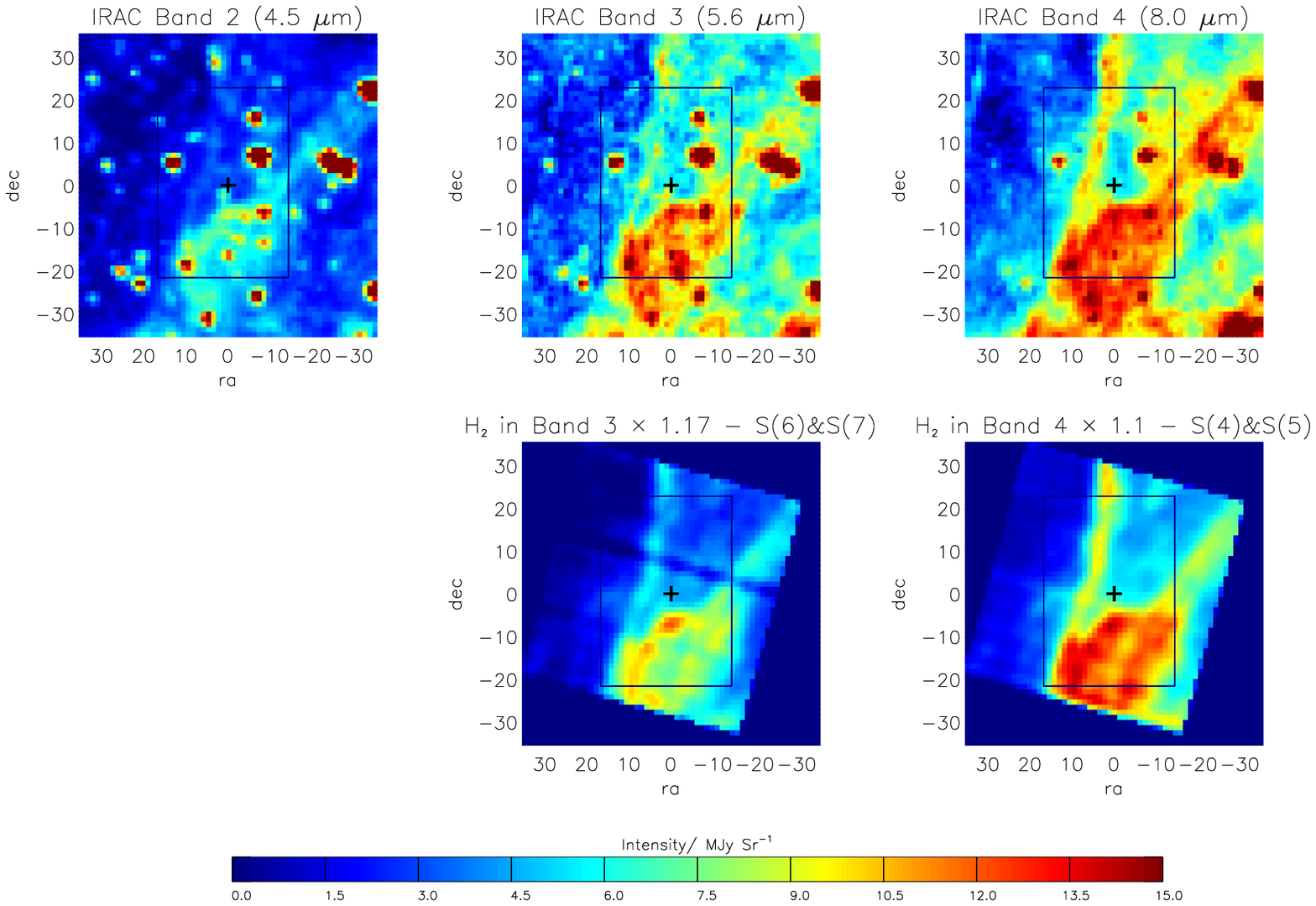}
\caption{Similar as figure 1 but for W44. The (0,0) position is now
at $\alpha = 18^{h}56^{m}28^{s}.4$ , $\delta= 01^{\circ}29'59''$
(J2000). The center of the $\it{ISO}$ observation (both SWS and LWS)
coincides with the (0, 0) position in the
maps.}\label{fig:irs&iracW44}
\end{figure}

\begin{figure}
\includegraphics[scale=0.8]{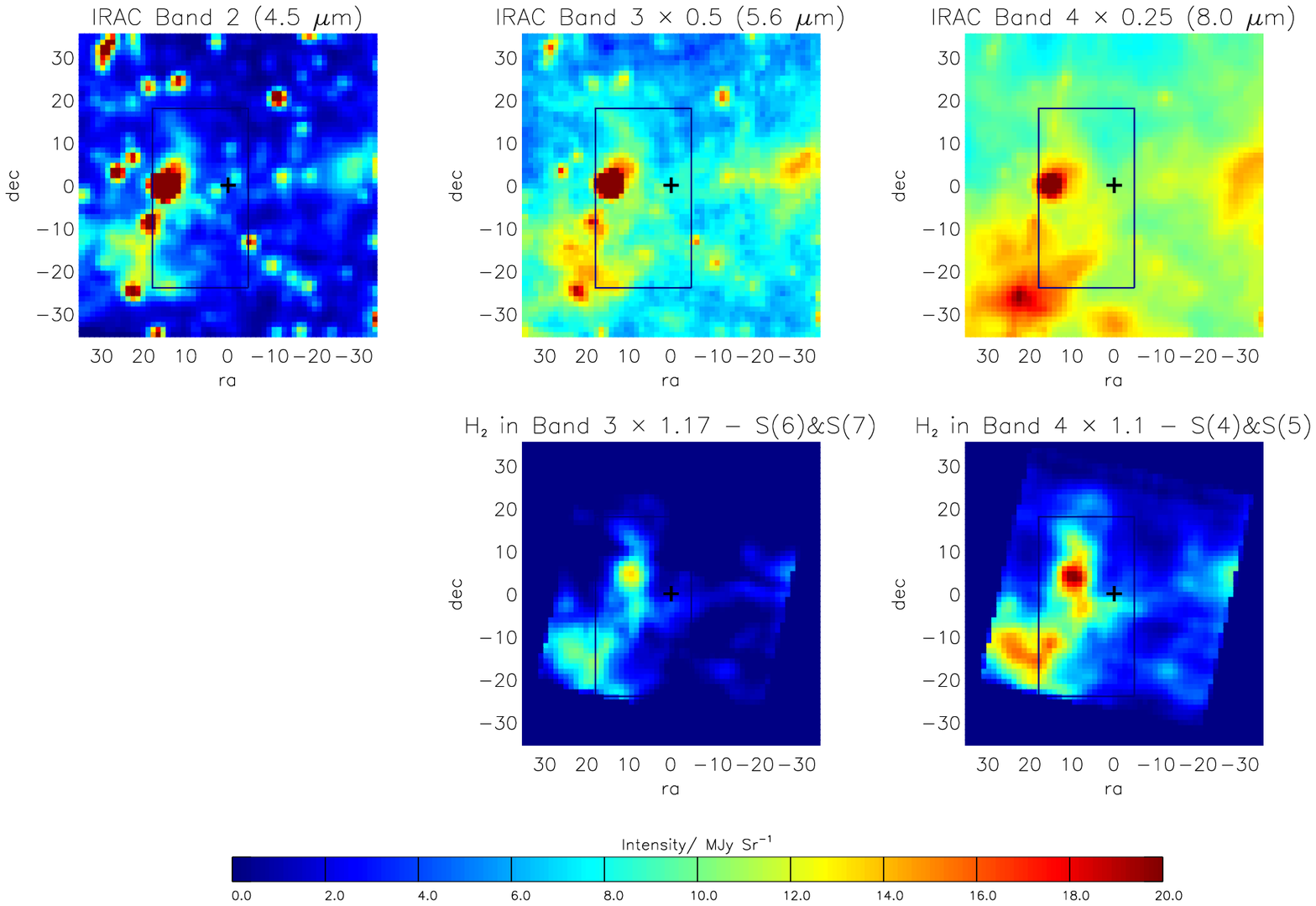}
\caption{Similar as figure 1 but for 3C391. The (0,0) position is
now at $\alpha = 18^{h}49^{m}21.^{s}.9$, $\delta=
-0^{\circ}57'22''$ (J2000). The center of the $\it{ISO}$
observation(both SWS and LWS) coincides with the (0, 0) position in
the maps.}\label{fig:irs&irac3C391}
\end{figure}

In the {\it Spitzer} archive, two W28 IRAC maps in Program 20201,
two W44 IRAC maps and two 3C391 IRAC maps in Program 186 are found
to cover the same regions observed by IRS. IRAC maps for the same
source are averaged to get better signal--to--noise ratio for each
pixel. The comparison between the IRAC maps and the H$_2$
contributions for W28, W44 and 3C391 are shown in Figure
\ref{fig:irs&iracW28} , \ref{fig:irs&iracW44} and
\ref{fig:irs&irac3C391}, respectively.  The IRS spectra, averaged
over a Gaussian beam with HPBW $\sim$ 25" for each SNR source, are
presented in N07. For all three SNRs, the PAH features are
moderately strong within the IRAC 5.8$\mu$m and 8$\mu$m bands.
Furthermore, all the IRAC maps are heavily polluted by point
sources. For 3C391, the IRS map shows bright [Fe II] 5.34 $\mu$m
emission from the western knot.

\begin{figure}
\includegraphics[scale=0.8]{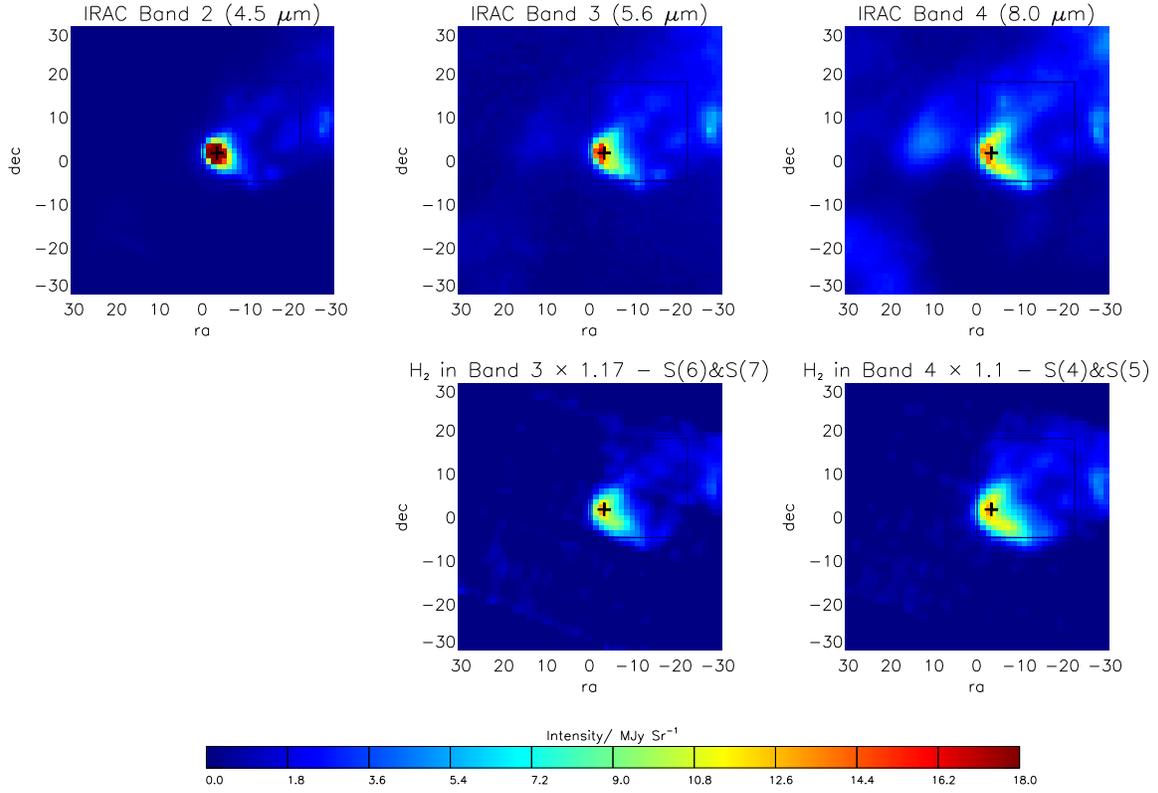}
\caption{Similar as figure 1 but for HH7. The (0,0) position is now
at $\alpha = 3^{h}29^{m}8^{s}.6$ , $\delta= 31^{\circ}15'26''.8$
(J2000). The center of the 80$''$ LWS beam observing the CO lines is
marked by the crosses.}\label{fig:irs&iracHH7}
\end{figure}

 The IRS spectra of HH7 and HH54 are shown in N06, which gives
average spectra for 15" diameter circular apertures. We have
extracted three HH7 IRAC maps obtained in Programs 6, 178 and 30516
from the $\it Spitzer$ archive. A comparison between the two sets of
maps are shown in Figure \ref{fig:irs&iracHH7}. The IRAC band 4
(8$\mu$m) intensity for HH7 appears to come mainly from H$_{2}$ S(4)
and S(5) line emissions; by contrast, the peak intensity in IRAC
band 3 (5.8$\mu$m) is 50$\%$ stronger than that expected from the
corresponding H$_{2}$ maps, which implies the presence of additional
contributions from dust continuum emissions. The extraordinarily
bright HH7 band 2 (4.5$\mu$m) intensity (compared with the
brightness of band 3 and band 4) implies that it is probably
dominated by continuum emission rather than H$_{2}$ emissions.

\begin{figure}
\includegraphics[scale=0.8]{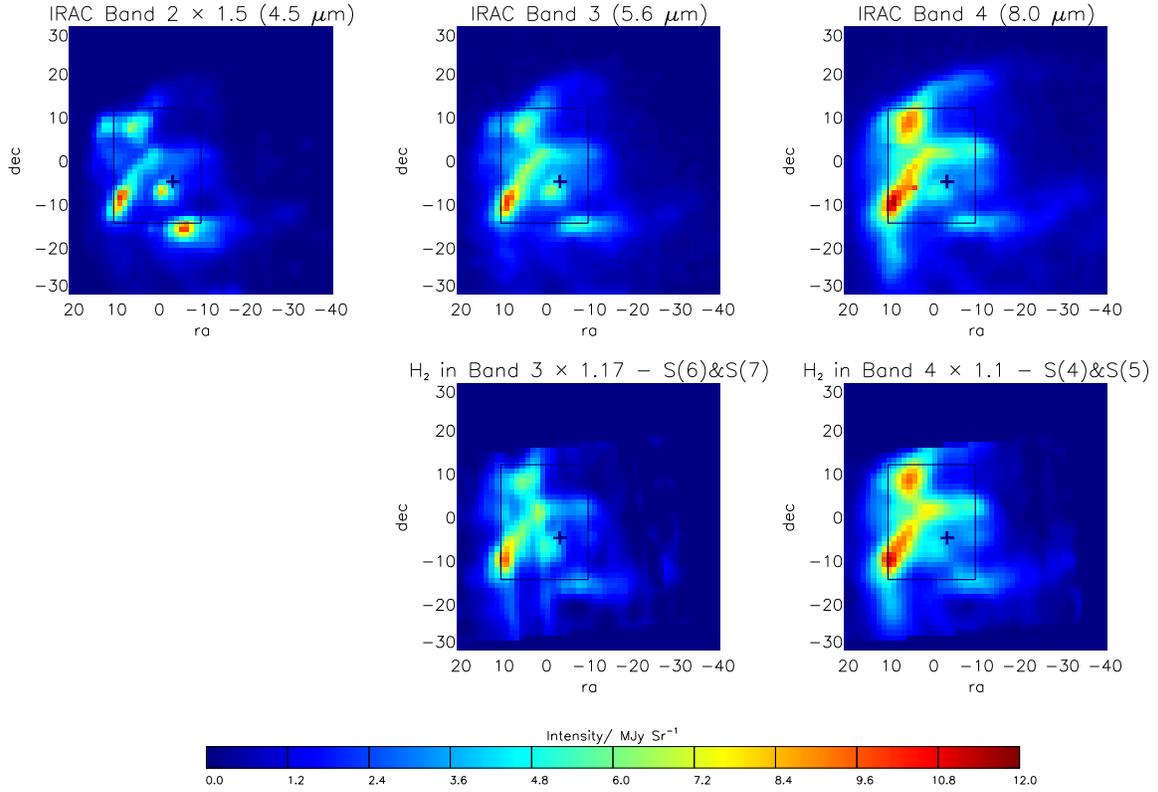}
\caption{Similar as figure 1 but for HH54. The (0,0) position is now
at $\alpha = 12^{h}55^{m}51^{s}.5$ , $\delta= -76^{\circ}56'19''.1$
(J2000). The center of the 40$''$ LWS beam observing the CO line
emissions is marked by the crosses.}\label{fig:irs&iracHH54}
\end{figure}

Two HH54 IRAC maps were obtained in Program 176. The similarity of
the IRAC and IRS H$_2$ spectral images are apparent; see Figure
\ref{fig:irs&iracHH54}. We have applied the same analysis method as
in NY08 and found that, as for IC443C, the two IRAC bands for HH54
(5.8$\mu$m and 8$\mu$m) are mainly accounted for by H$_2$ emissions.
It should be noted that N06 detected a weak [Fe II] fine structure
line at 5.34 $\mu$m toward HH54, which should contribute less than
18$\%$ of the 5.8 $\mu$m band intensity for most positions in the
map. The morphology of the HH54 4.5 $\mu$m band emission is similar
to that of the IRS H$_2$ spectral line maps, though it is a little
more clumpy, implying that the H$_2$ line emissions are also very
important components within band 2. We are assuming here that for
HH54, like IC443C, the IRAC 4.5 $\mu$m band (band 2) is also
dominated by H$_2$ emissions, i.e. the contribution from other
species is less than 25$\%$ in band 2, which is the typical error in
the line intensity. Other possible contributors include atomic
hydrogen recombination lines -- Br$\alpha$ and Pf$\beta$ -- which
should be weak for molecular shocks, [Fe II] fine structure lines,
and dust continuum emission. The possibility of strong CO v = 1 -- 0
emission in the 4.5 $\mu$m band was shown to be unlikely by NY08.
NY08 considered collisional excitation of CO v = 1 -- 0 transitions
by H and H$_2$ and found that even at a H/H$_2$ ratio $\sim 1$ and a
high density with $n$(H$_2)=10^{6}$ cm$^{-3}$, the fractional
contribution of CO emission to the 4.5 $\mu$m band is less than
$20\%$. For gas with H/H$_2$ $\sim 10\%$ and $n$(H$_2)=10^{4}$
cm$^{-3}$, the fraction would be less than $2\%$.

\subsection{Additional constraints imposed by {\it ISO} observations of CO and H$_{2}$}

The two complementary spectrometers on {\it ISO} provide us with
more molecular data for these shock-excited regions. The
Long-Wavelength Spectrometer (LWS) is designed for spectroscopic
observations in the range of 43 -- 196.9 $\mu$m and covers many
high-lying CO pure rotational transitions. The detection of CO
emission with $J$ $>$ 14 in LWS observations has been reported for
all six sources.  Snell et al.\ (2005) presented observations of
three CO lines, $J$ = 15 -- 14, $J$ = 16 -- 15, $J$ = 17 -- 16,
within the 80$''$ LWS beam centered on IC443C from the $\it{ISO}$
archive. In the LWS observations carried out by Reach \& Rho (1998),
four CO lines were detected toward 3C391, viz.\ $J$ = 14 -- 13, $J$
= 15 -- 14, $J$ = 16 -- 15 and $J$ = 17 -- 16. For W28, they
detected only the CO $J$ = 15 -- 14 and $J$ = 16 -- 15 transitions.
For W44, only the $J$ = 16 -- 15 line could be identified. Molinari
et al.\ (2000) observed several locations along the HH 7 -- 11 flow
and detected four CO rotational lines within the LWS beam on HH7:
$J$= 14 -- 13, $J$ = 15 -- 14, $J$ = 16 -- 15 and $J$ = 17 -- 16.
The LWS CO spectrum for HH54 was taken toward a position called
HH54B, marked by the crosses in Figure 6. Six CO rotational
features, $J$ = 14 -- 13, $J$ = 15 -- 14, $J$ = 16 -- 15, $J$ = 17
-- 16, $J$ = 18 -- 17 and $J$ = 19 -- 18, were reported by Nisini et
al.\ (1996) and Liseau et al.\ (1996).  With an improved Relative
Spectral Response Function, Giannini et al.\ (2006) presented a new
analysis of the spectra, leading to the detection of CO $J$ = 20 --
19. For these observations, the measured line fluxes had calibration
uncertainties estimated to be up to $\sim$ 30$\%$.  With larger
critical densities of the order of 10$^{7}$ cm$^{3}$ , these CO
high-lying rotational lines can provide sensitive diagnostics for
probing density in these regions.

Besides the CO emissions, Reach and
Rho (2001) obtained spectra of the H$_2$ S(9) and S(3) lines for
W28, W44 and 3C391 within the $14''\times20''$ aperture of the Short
Wavelength Spectrometer (SWS) on $\it{ISO}$. The central positions of
all these $\it{ISO}$ observations are marked by the crosses on
Figures 1 -- 6. For W28, W44 and 3C391 the LWS and SWS observations
share the same beam centers and are consistent with the
$(0,0)$ positions of the IRS maps. The measured H$_2$ S(9)/S(3)
ratios are another valuable diagnostic tool and are used in our
fits to constrain the best-fit parameters of the gas. The LWS
measured CO line fluxes along with the 1 $\sigma$ errors and the SWS
S(9)/S(3) ratios are all listed in Table 1.

\section{Molecular Emission from C-type shocks}

\subsection{The Excitation Model}

The H$_{2}$ emission spectrum is among the most important
diagnostics needed to constrain conditions in shocked molecular gas
as well as to distinguish between different shock models. Over the
last several decades, it has been widely observed that the
rotational diagrams of H$_{2}$ often exhibit positive curvatures.
This kind of curvature can not be accounted for by extinction
effects only, which affect the H$_{2}$ S(3) line much more strongly
than the other rotational lines, and may imply the existence of a
mixture of gas temperatures. N06 and N07 investigated the H$_{2}$
excitation diagrams for the six sources we are studying here, in
which the molecular hydrogen emission was modeled with a combination
of gas at two temperatures. In this paper, we adopt a power-law
temperature distribution similar to that described by NY08, with the
column density of gas at temperature between $T$ and $T$ + $dT$
assumed to be proportional to $T^{-b}$. Instead of the lower
temperature limit $T_{min}$ = 300 K adopted by NY08, we extend it to
100 K here because warm gas at 100 -- 300 K can contribute
significantly to those low-lying H$_{2}$ emissions accessible to
IRS, especially for v = 0 -- 0 S(0) and S(1). This power-law
distribution is consistent with the prediction from the bow-shaped
C-shock model developed by Smith, Brand \& Moorhouse (1991) (NY08).
Smith, Brand \& Moorhouse found that bow shocks can produce gas at a
wide range of excitation temperatures and thus provide a way of
explaining the H$_{2}$ line ratios observed for many sources, while
planar shock models fail to reproduce the observed ratios.

Another interesting characteristic of the H$_{2}$ rotational
diagrams lies in the zigzag pattern, corresponding to
non-equilibrium H$_{2}$ ortho-to-para ratios. This phenomenon is
especially notable for the five sources W28, W44, 3C391, HH7 and
HH54 (N06; N07). With a closer look, it is quickly apparent that the
zigzag tends to ``diminish'' for high--lying levels. Given the fact
that low- and high-excitation lines are produced by different
temperature components, N06 argued that the change in the degree of zigzag
is caused by the strong temperature dependence of the
para-to-ortho conversion efficiency. With this process dominated by
collisions with atomic hydrogen in C-type shocks (Timmermann 1998;
Wilgenbus et al., 2000), the current OPR is given by equation (1) in
Neufeld et al. (2009; hereafter N09), as a function of initial ratio
OPR$_{0}$, atomic hydrogen density n(H) and shock-heating period
$\tau$. These three values along with the number and column density
of H$_{2}$ | $n$(H$_{2}$) and $N$(H$_{2}$) |
determine the H$_{2}$ emission spectrum.

\subsection{Constraining the physical parameters}

For an excitation model with the parameters discussed above, the molecular
line intensities are easy to calculate if statistical equilibrium is
achieved, which is an approximation widely adopted for modeling
molecular emission in shocks. We have confirmed the validity of this
approximation for transitions of the three species H$_{2}$, HD and
CO accessible in the infrared observations mentioned in this paper
(see Appendix A). The rate equations are then simplified as
\begin{equation}
\sum_{J'}\left (C_{J'\rightarrow J}+A_{J'\rightarrow J}\right
)f_{J'}-f_J \cdot \sum_{J'}\left (C_{J\rightarrow
J'}+A_{J\rightarrow J'}\right)=0,
\end{equation}
where $f_J$ represents the fractional population in state $J$, $A_{J'\rightarrow J}$
is the rate of spontaneous decay from state $J'$ to $J$, and $C_{J'\rightarrow J}$
is the rate of collisionally-induced transitions from $J'$ to $J$.  We
consider collisional excitation (de-excitation) and spontaneous
decay processes only, and for H$_2$ and HD we neglected optical depth effects.
For the H$_2$ collisional rate coefficients, we
used data computed by Flower \& Roueff (1999), Flower et
al.\ (1998), Flower \& Roueff (1998b) and Forrey et al.\ (1997). The
rate coefficients for HD were adopted from Flower (1999), Roueff \&
Zeippen (1999) and Roueff \& Flower (1999). For CO, we made use of
collisional data from Flower (2001), Wernli et al.\ (2006) and
Balakrishnan et al.\ (2002), together with the extrapolation presented
by Schoier et al.\ (2005) for temperatures up to 2000 K
and CO rotational states up to $J = 40$. For cases where optical depth
effects may become non-negligible, specifically for the CO transitions,
we applied the large-velocity-gradient (LVG) approximation and
multiplied the radiative transition part in equation (1) by a term
$\beta$, which is called the escape probability for photons. Neufeld
\& Kaufman (1993) suggested an angle-averaged probability for planar
shocks
\begin{equation}
\beta={1\over 1+3\tau_s}
\end{equation}
where $\tau_s$ is the Sobolev optical depth. An average velocity
gradient of 2 $\times 10^{-10}$ cm s$^{-1}$/cm and a
$n$(CO)/$n$(H$_2$) abundance ratio of 5 $\times 10^{-5}$ are adopted in calculating
the CO optical depth.  We note however, that the optical thickness
of CO affects these highly excited rotational states measured in LWS
observations by less than 10\%.

We have adopted two approaches to find the best-fit parameters in
the shock model using a $\chi^{2}$ minimization method.  In the first approach,
we considered only the
H$_{2}$ spectral lines S(0) -- S(7) observed by IRS; while in the second
approach we fitted all the
available data, including the IRS H$_{2}$ intensities, the IRAC band 2
intensities -- if dominated by high-lying H$_{2}$ rotational transitions
(IC443C and HH54) -- the HD R(3) \& R(4) intensities, and the CO line fluxes
obtained from the $\it{ISO}$
archive.  We decided not to use the IRAC 3.6 $\mu$m band intensities, which are
attributable to H$_{2}$ ro-vibrational emissions. The excitation of
H$_{2}$ vibrational transitions, unlike the pure rotational lines
we considered, is dominated by collisions by atomic hydrogen even
with a small $n$(H)/$n$(H$_{2}$) ratio $\sim$ 3$\%$ (NY08). Since
the collisional dissociation processes for molecular hydrogen are
more efficient within the hotter component of the shocked gas, the dependence of
the H$_{2}$ vibrational emission upon temperature will be more
complicated than what our model describes. Thus, we exclude IRAC
band 1 (3.6 $\mu$m) in the fitting to avoid it affecting our
estimate of the best fit temperature distribution and density. For
the other transitions mentioned above, the excitation processes should
be always dominated by collisions with H$_{2}$ given the conditions
in molecular shocks.  In our calculation, we take into account
collisional excitation by H$_{2}$ and helium.  A helium
abundance $n$(He)/$n$(H$_{2}$) of 0.2 is assumed. Errors introduced
by neglecting collision with atomic hydrogen are largest for the CO
lines and those high-lying H$_{2}$ transitions contributing to IRAC
band 2, which are however still less than 15$\%$ given an average
$n$(H)/$n$(H$_{2}$) ratio of 10$\%$ , which is the upper limit of the
value for IC443C estimated by Burton et al.\
(1988) based on the Br$\gamma$ intensity.

\section{Results}

As mentioned in Section 3, we used two ways to derive the best-fit
parameters. In the first approach, only IRS H$_2$ lines were
considered. In the second approach, all available molecular data were
included in the fitting process. In the case of IC443C and HH54, these data
comprise the H$_{2}$ \& HD line intensities observed by IRS, the IRAC band 2 intensity,
and the CO line intensities observed by {\it ISO}/LWS. For W28,
W44 and 3C391, we included the H$_{2}$ lines and SWS-measured
S(9)/S(3) ratios. For HH7, only the IRS-observed H$_{2}$ and HD lines were used.
We found that the uncertainties in the LWS-measured CO line intensities for
W28, W44, 3C391 and HH7 are too large to provide useful information
about the physical conditions in the gas.

To deredden the IC443C line fluxes, we originally tried the
extinction correction with $A_{2.12\mu m}$ = 1.3 -- 1.6 by Richter
et al.\ (1995) and the R$_{V}$ = 3.1 extinction curves from
Weingartner \& Draine (2001), which ended up with a S(3) intensity
almost two times larger than expected given the other H$_2$ line
intensities. Treating the extinction as a free parameter in our fit
to the H$_2$ line intensities yielded an A$_V$ close to zero.  Thus,
we applied no extinction correction for IC443C in the following
calculations.  For W28, we applied an extinction correction with
E$_{B - V}$= 1 -- 1.3 given by Long et al.\ (1991) derived from the
[S II] line ratios. This value is also consistent with the estimate
by Bohigas et al.\ (1983) who obtained E$_{B - V}$= 1.16. For W44,
the absorbing column density along the line-of-sight to this region
is estimated to be $\sim 2\times10^{22}~$cm$^{-2}$ (Rho et al.\
1994), corresponding to an A$_V$ $\sim$ 10. For 3C391, Reach et al.\
(2002) suggested a visual extinction of A$_V$ = 19 for the
IRS-observed region, which is denser than other parts of the cloud,
based upon an upper limit on the foreground column density of
$(2-3.6) \times 10^{22}$ cm$^{-2}$ inferred from a spectral analysis
of the X-ray data (Rho $\&$ Petre 1996). For HH7, we adopted E$_{J -
K}$ = 0.7, as estimated by Gredel (1996) for the neighboring source
HH 8, and for HH54 we assumed A$_V= 1.64$, following Gredel (1994).

\begin{figure}
\includegraphics[scale=0.7]{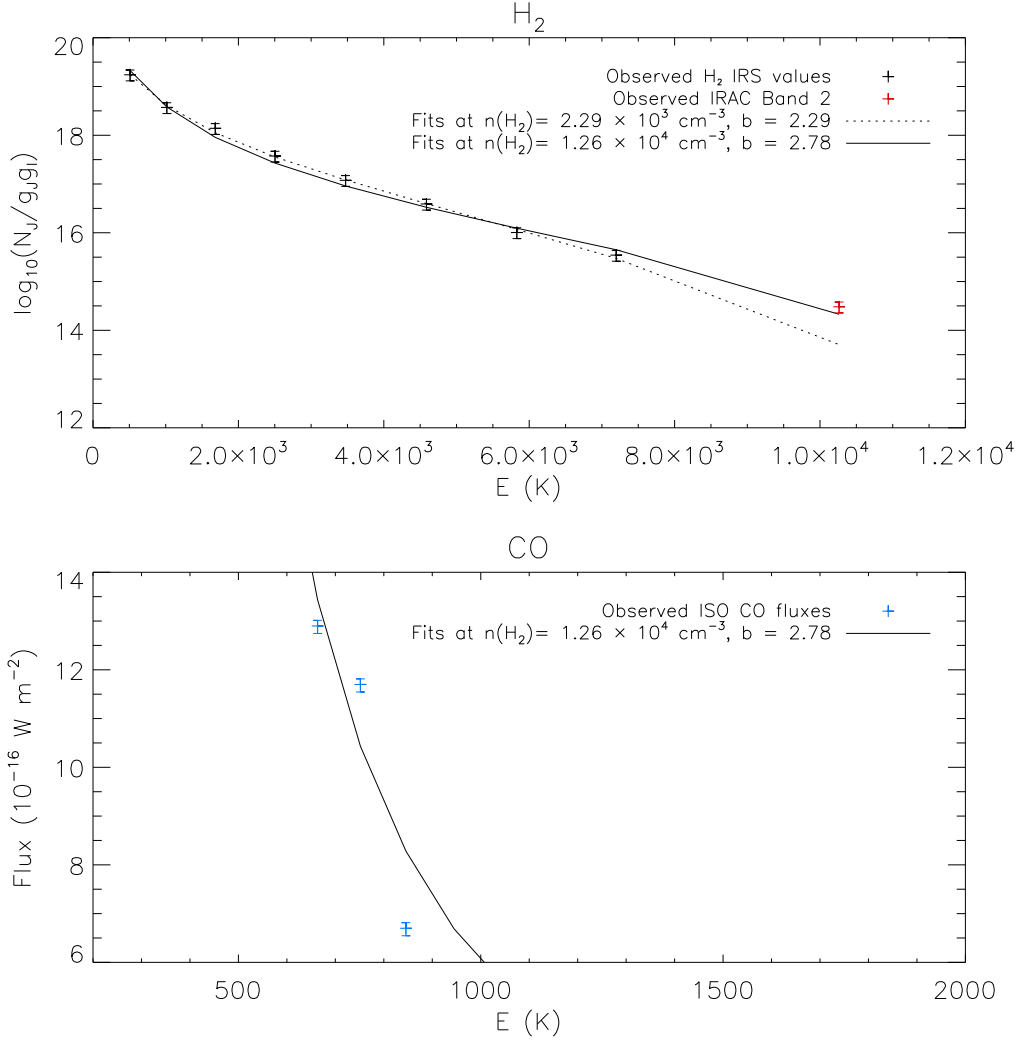}
\caption{The upper panel shows best fits to IC443C H$_2$ rotational
diagram and the lower panel represents fits to the CO excitation
diagram. We adopted two different approaches in the fitting. The
dotted line represents fits computed with the IRS H$_2$ lines S(0)
to S(7) only and solid lines represent fits to all features
including IRS H$_2$ and HD lines, IRAC band 2 (4.5 $\mu$m) emission,
LWS CO fluxes $J$ = 15--14, $J$ = 16--15 and $J$ = 17--16.  To show
fit to IRAC band 2 (4.5 $\mu$m) , we assume that half of the band 2
intensity comes from H$_2$ S(9) and plot that false ``S(9)" on the
H$_2$ rotational diagram, represented by the red cross on the upper
panel. The error bars for each line are plotted assuming 30\%
uncertainties for LWS CO fluxes and 25\% for the
rest.}\label{fig:rotIC443}
\end{figure}

\begin{figure}
\includegraphics[scale=0.7]{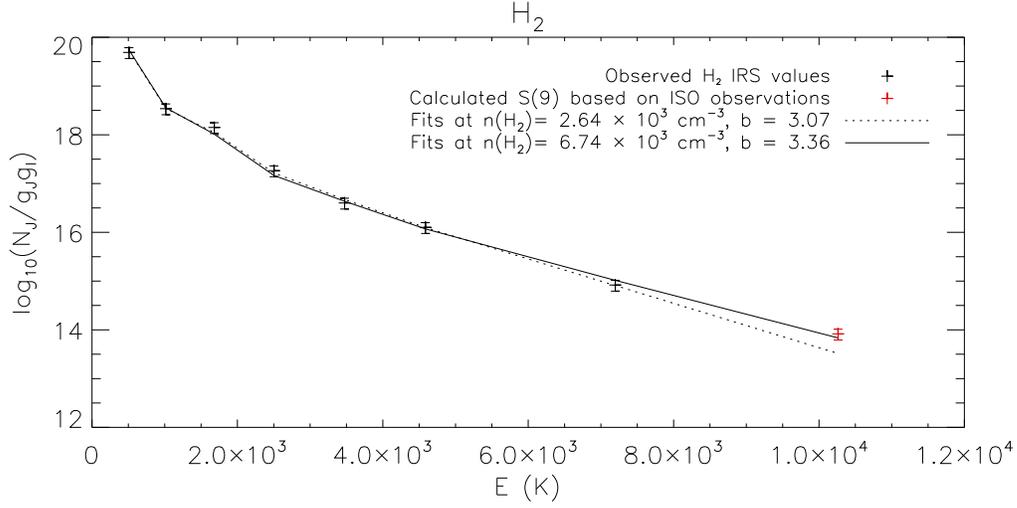}
\caption{Similar to Figure \ref{fig:rotIC443}. The dotted line
represents fits to W28 H$_2$ rotational diagram computed with the
IRS H$_2$ lines. The S(6) line flux is excluded which can not be
measured reliably. The solid line shows fits calculated with both
IRS H$_2$ lines and S(9)/S(3) ratio obtained by the $\it{ISO}$ SWS
observation. The rightmost red cross shows S(9) evaluated with IRS
S(3) intensity and the given SWS S(9)/S(3) ratio.
}\label{fig:rotW28}
\end{figure}

\begin{figure}
\includegraphics[scale=0.7]{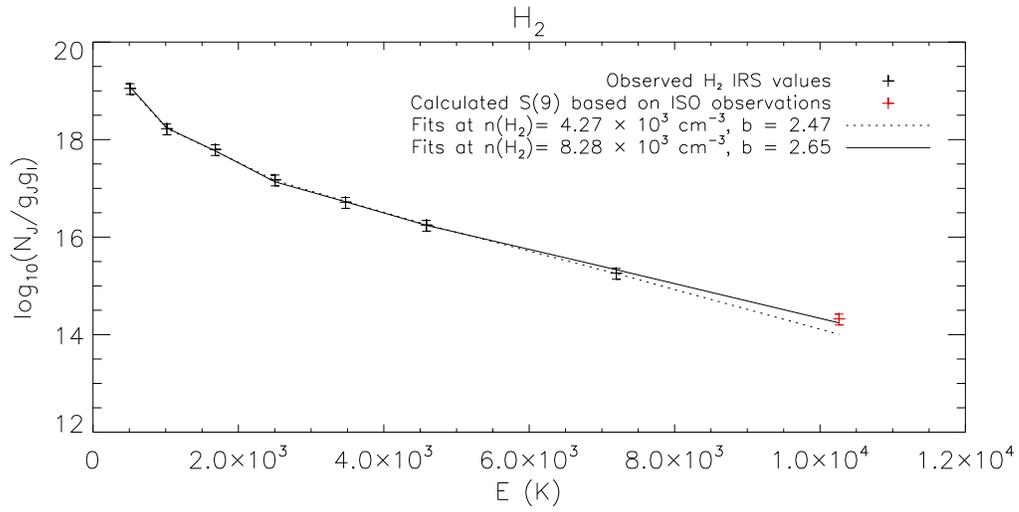}
\caption{Same as Figure \ref{fig:rotW28} but for W44. The H$_2$ S(6)
line is not included as it cannot be measured reliably.
}\label{fig:rotW44}
\end{figure}

\begin{figure}
\includegraphics[scale=0.7]{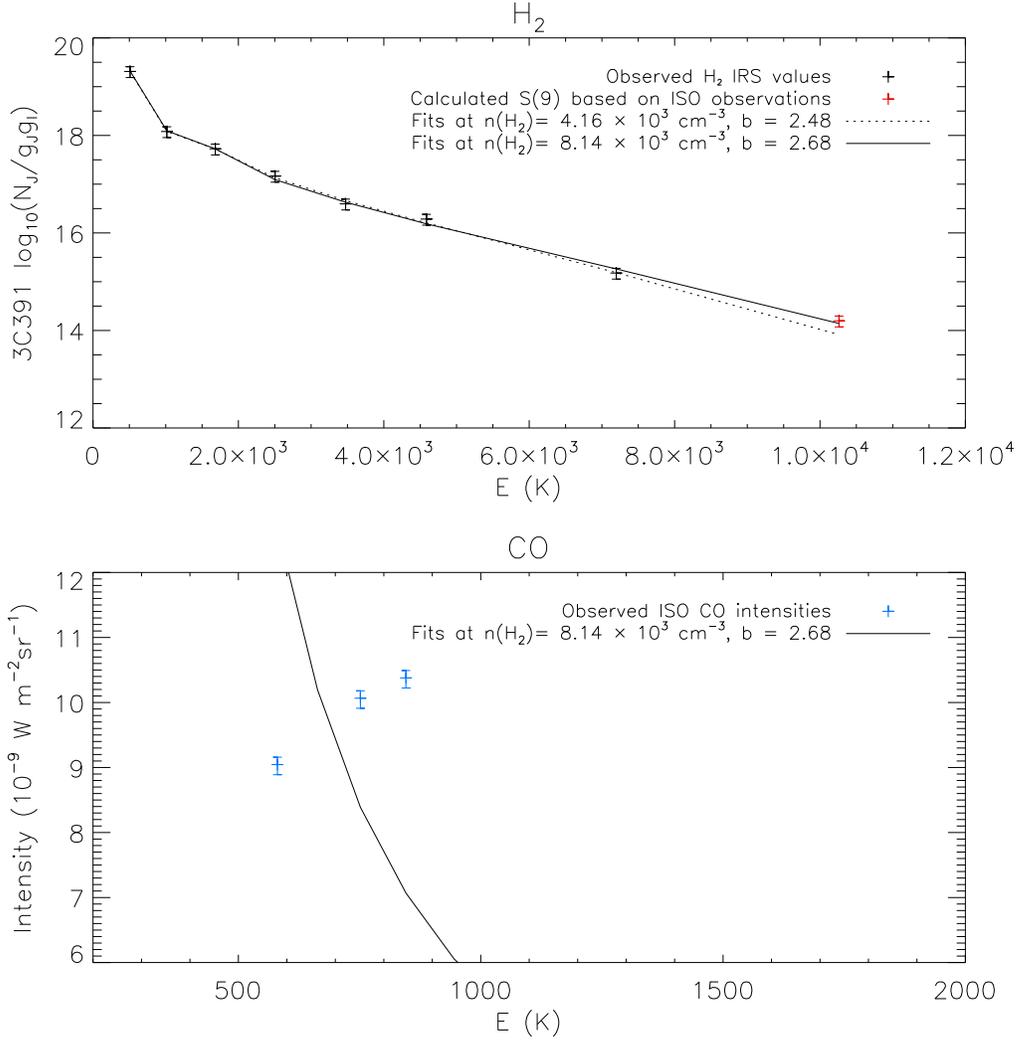}
\caption{Same as Figure \ref{fig:rotIC443} but for 3C391. The H$_2$
S(6) line is not included as it cannot be measured reliably. Though
fits to the CO rotational diagram is still shown in the lower panel,
these CO fluxes observed in LWS observation, including $J$ = 14--13,
$J =$ 15--14, $J =$ 16--15 and $J =$ 17--16, are excluded in our
computation because of large uncertainties.  Noted here the observed
CO flux for $J =$ 15--14 is way above the range of the plot, which
may result from unreliable measurement. }\label{fig:rot3C391}
\end{figure}

\begin{figure}
\includegraphics[scale=0.7]{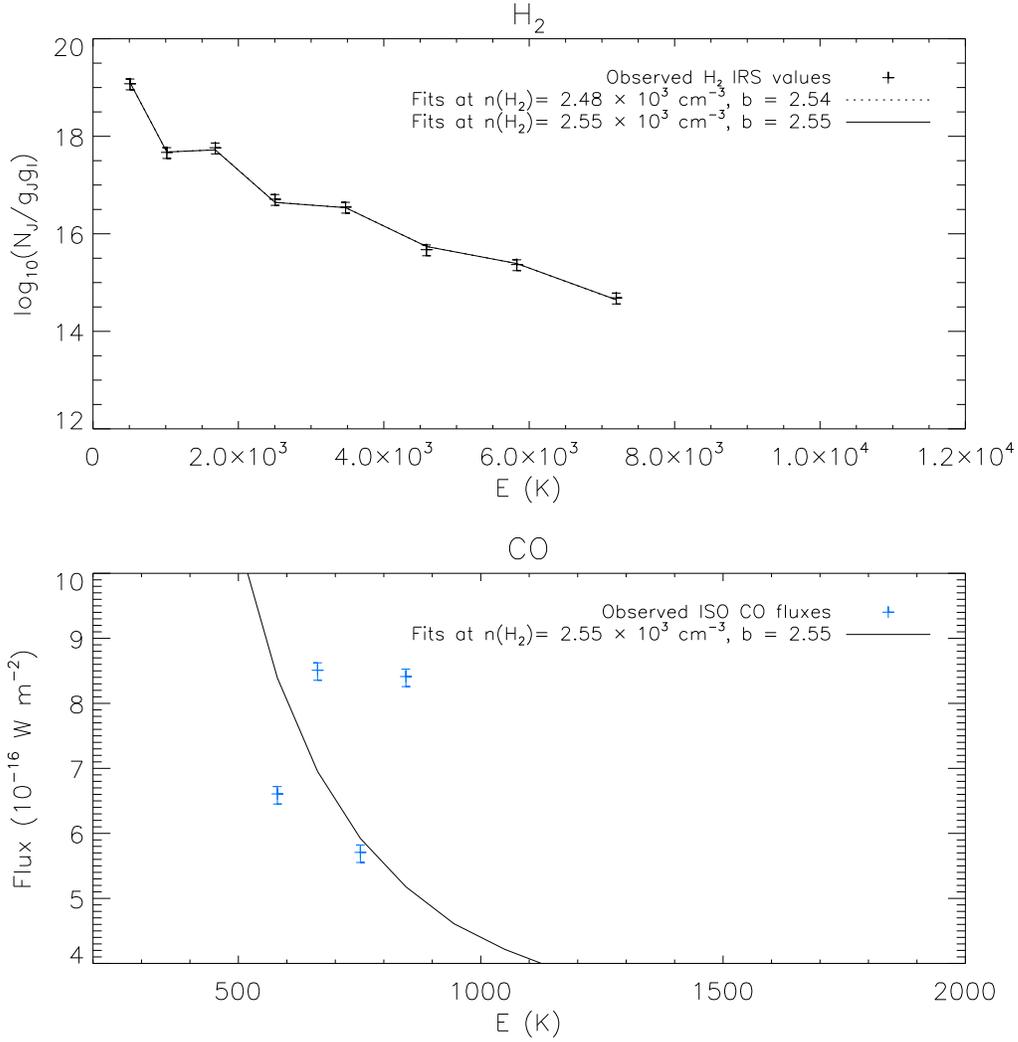}
\caption{Same as Figure \ref{fig:rotIC443} but for HH7. Solid lines
represent fits to IRS H$_2$ and HD lines. Since the two fits are
very close the dotted and solid lines overlap. Though fits to the CO
rotational diagram is still shown in the lower panel, the CO fluxes
observed in LWS observation, including $J = 14$--13, $J =$ 15--14,
$J = 16$--15 and $J = 17$--16, are excluded in our computation due
to large uncertainties. }\label{fig:rotHH7}
\end{figure}

\begin{figure}
\includegraphics[scale=0.7]{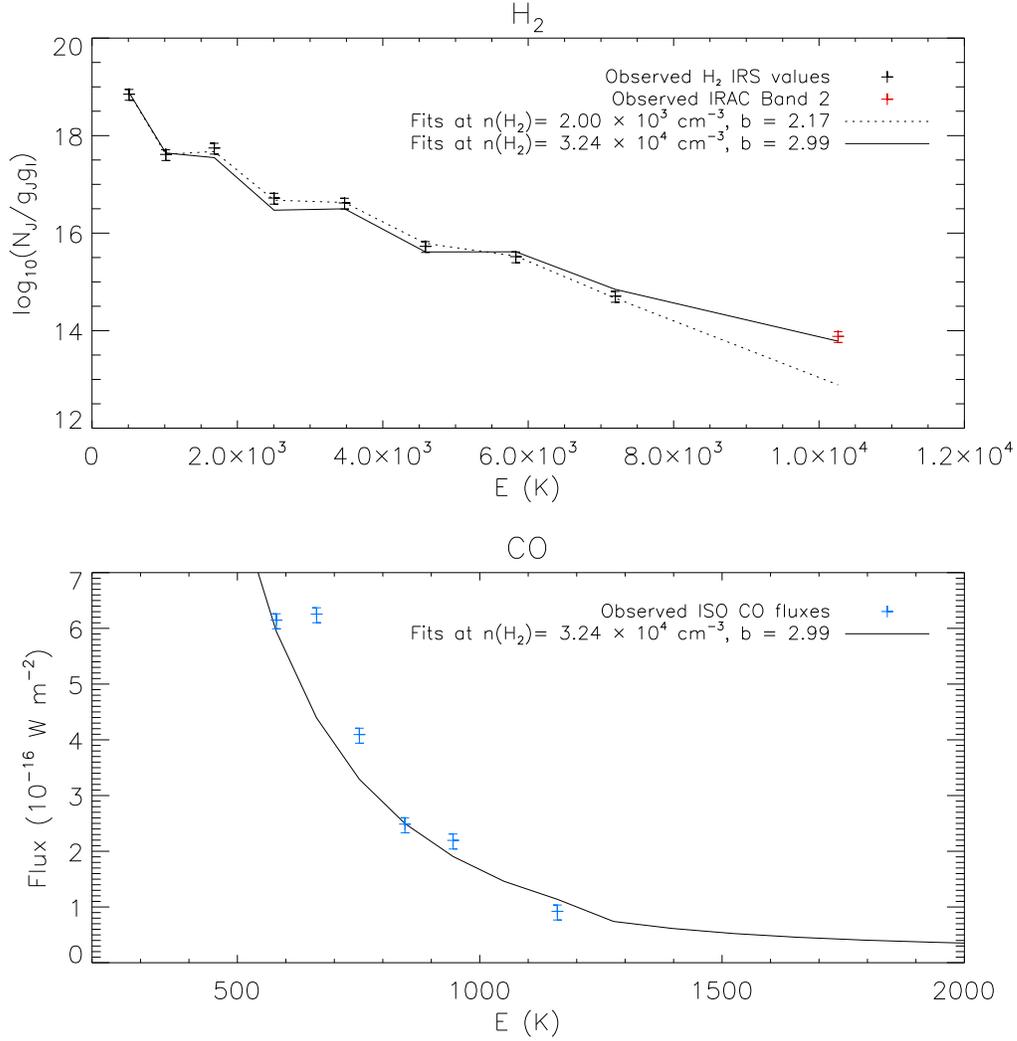}
\caption{Same as Figure \ref{fig:rotIC443} but for HH54. The blue
crosses in the lower panel represent observed CO fluxes at $J =
14$--13, $J = 15$--14, $J = 16$--15, $J = 17$--16, $J = 18$--17 and
$J = 20$--19 respectively.}\label{fig:rotHH54}
\end{figure}

The best fits to the H$_{2}$ and CO rotational diagrams are shown by
the dotted lines (first approach) and solid lines (second approach)
in Figures \ref{fig:rotIC443} to \ref{fig:rotHH54}. In the case of
H$_2$ S(9), we assumed an S(9)/S(3) line ratio equal to that
measured by {\it ISO}/SWS for the sources W28, W44 and 3C391. For
IC443C and HH54, we estimated the S(9) line intensity from the IRAC
band 2 intensity, assuming one-half of the emission in that band to
result from H$_2$ S(9), roughly consistent with the fractional
contribution obtained by NY08 ($62\%$ for one specific excitation
model with an assumed $b$ of 4.5 and $n$(H$_2)$ of $10^{6}~$
cm$^{-3}$ reported in their Table 1). Though the CO lines are
excluded in the fits for 3C391 and HH7, the excitation diagrams of
CO for these two sources were still presented in Figures
\ref{fig:rot3C391} and  \ref{fig:rotHH7}.  The best-fit parameters
for each source are listed in Table 2.  From this table, we can see
that while the best-fit density determined from the H$_{2}$ pure
rotational transitions alone is $\sim 2 - 4\times10^{3}~$cm$^{-3}$,
a value several times larger provides the best-fit to the complete
data set including the H$_2$ S(9), IRAC band 2, HD and CO line
intensities.

Although the IRS maps are smaller than the whole regions that
contribute to the 80$''$ LWS beam, we can place an upper limit on the
average CO abundance by comparing the H$_2$ line fluxes obtained in each entire
IRS map with the CO fluxes obtained with LWS.
This method yields the firm upper limits $n$(CO)/$n$(H$_2$) $< 1.5\times 10^{-3}$ and
$n$(CO)/$n$(H$_2$) $< 5.5\times10^{-4}$ for IC443C and HH54,
respectively.  If we assume the distribution of CO emission to be similar to
that measured in IRAC band 2, we obtain rough estimates of the CO abundance of
$\sim3-5\times 10^{-5}$ for IC443C and
$\sim2-4\times 10^{-5}$ for HH54.

\begin{figure}
\includegraphics[scale=0.7]{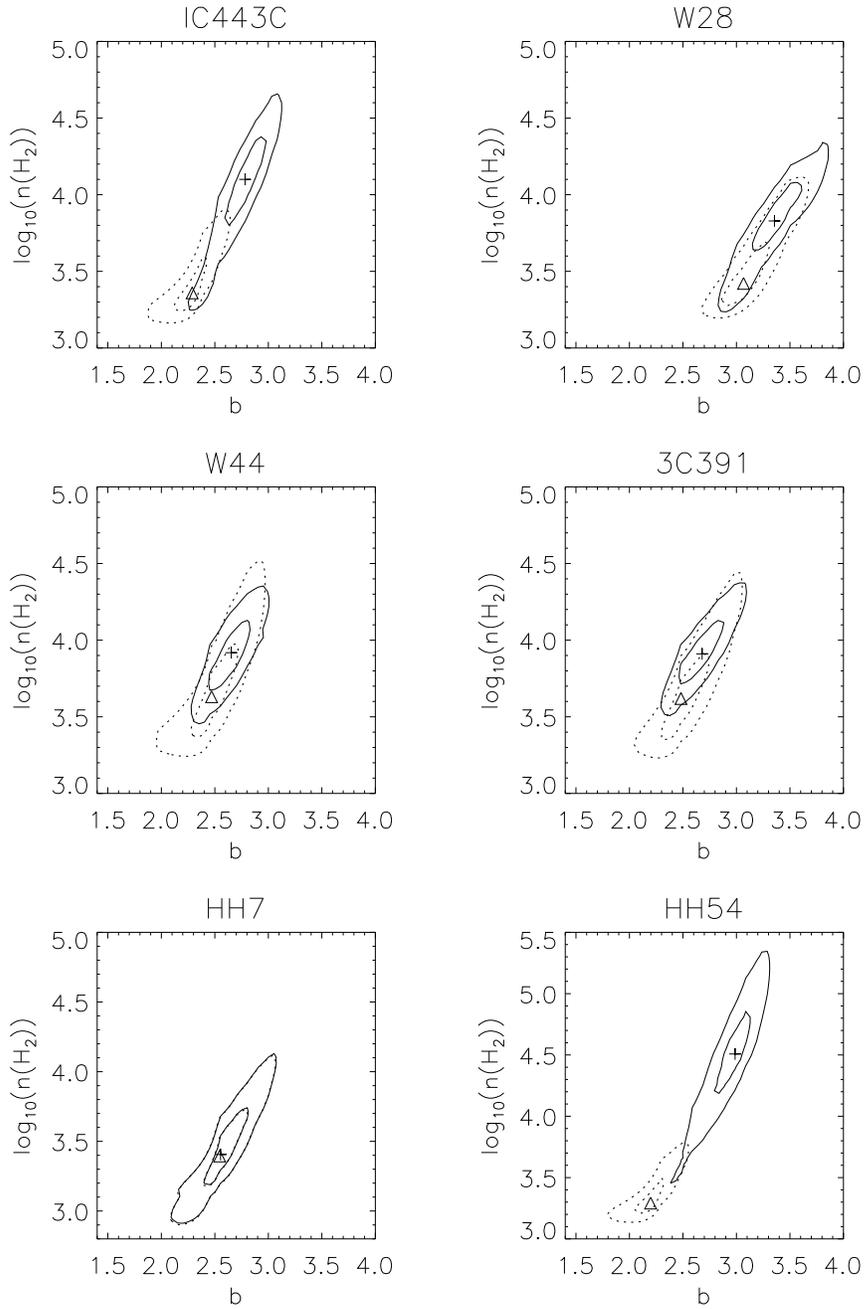}
\caption{$\chi^{2}$ contours in the $b$--n($H_{2}$) plane for all
six sources.  Dotted lines show fits calculated with IRS H$_2$ lines
and solid lines for fits with all reliable features included. The
best-fit $b$ and $n$(H$_{2}$) are marked by triangles and crosses
for the two cases respectively. The inner contours confine the
$68.3\%$ joint confidence interval and the outer contours confine
the $95.4\%$ confidence interval. For HH7 the two sets of contours
almost overlap each other. }\label{fig:lgnb}
\end{figure}

Errors in the fitted parameters are evaluated by plotting $\chi^{2}$
contours in multi-parameter spaces. The $\chi^{2}$ values are
computed assuming a fractional uncertainty of 30$\%$ for the CO line
fluxes and $25\%$ for all other line emissions. Figure
\ref{fig:lgnb} shows the $68.3\%$ and $95.4\%$ confidence regions in
the $b$ -- $n$(H$_{2}$) plane for all six sources, with dotted lines
for fits with IRS H$_2$ lines only and solid lines for fits with all
reliable data included (second approach discussed above). The
elongated shape of the $\chi^{2}$ contours in the $b$ --
$n$(H$_{2}$) plane arises because the two parameters are degenerate,
as mentioned in N09, such that increasing the density or decreasing
$b$ (which raises the fraction of hot gas) have a similar effect on
the excitation of mid- and high-lying transitions considered in the
calculation. For HH7, including the HD R(3) and R(4) lines in our
calculation does not significantly change the best-fit parameters.
The two sets of contour plots for HH7 almost overlap with each
other, as shown in Figure \ref{fig:lgnb}; this is because within the
banana-shaped region constrained by the H$_{2}$ emission, the HD
R(3) to R(4) ratio is not a sensitive function of the gas density or
the power-law index, $b$.

\begin{figure}
\includegraphics[scale=0.7]{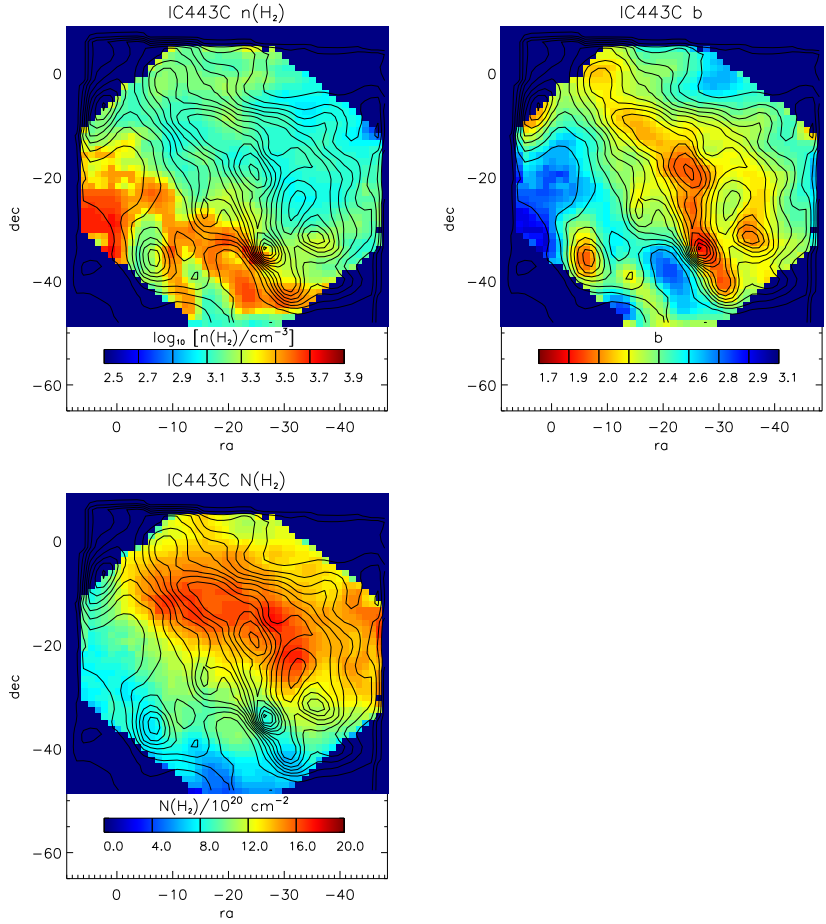}
\caption{Spatial distributions of the best-fit-parameters for
IC443C, including density $n$(H$_2$), power law index $b$ and column
density of gas $N$(H$_2$) with 100 K $<T<$ 5000 K, calculated with
IRS H$_2$ line emissions only. Regions are selected with S(5)
intensity larger than 2 $\times$10$^{-7}~$W m$^{-2}$sr$^{-1}$ so the
signal-to-noise ratios for each H$_2$ lines are good enough to yield
reliable fits. The contours of the brightest H$_2$ line S(5) are
superposed. }\label{fig:paraIC443}
\end{figure}

\begin{figure}
\includegraphics[scale=0.7]{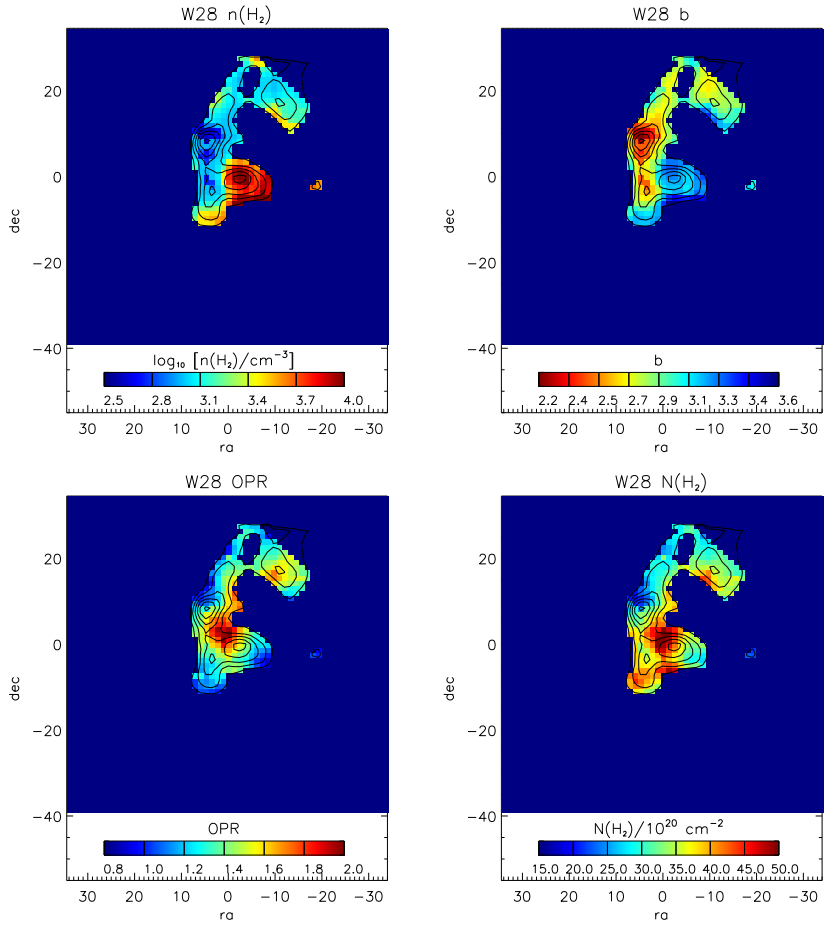}
\caption{Same as Figure \ref{fig:paraIC443} but for W28. Parameter
maps includes density $n$(H$_2$), power law index $b$, column
density of gas $N$(H$_2$) with 100 K $<T<$ 5000 K and averaged OPR
over the total column density. Regions are selected with S(5)
intensity larger than 7.5 $\times$10$^{-7}~$W
 m$^{-2}$sr$^{-1}$. }\label{fig:paraW28}
\end{figure}

\begin{figure}
\includegraphics[scale=0.7]{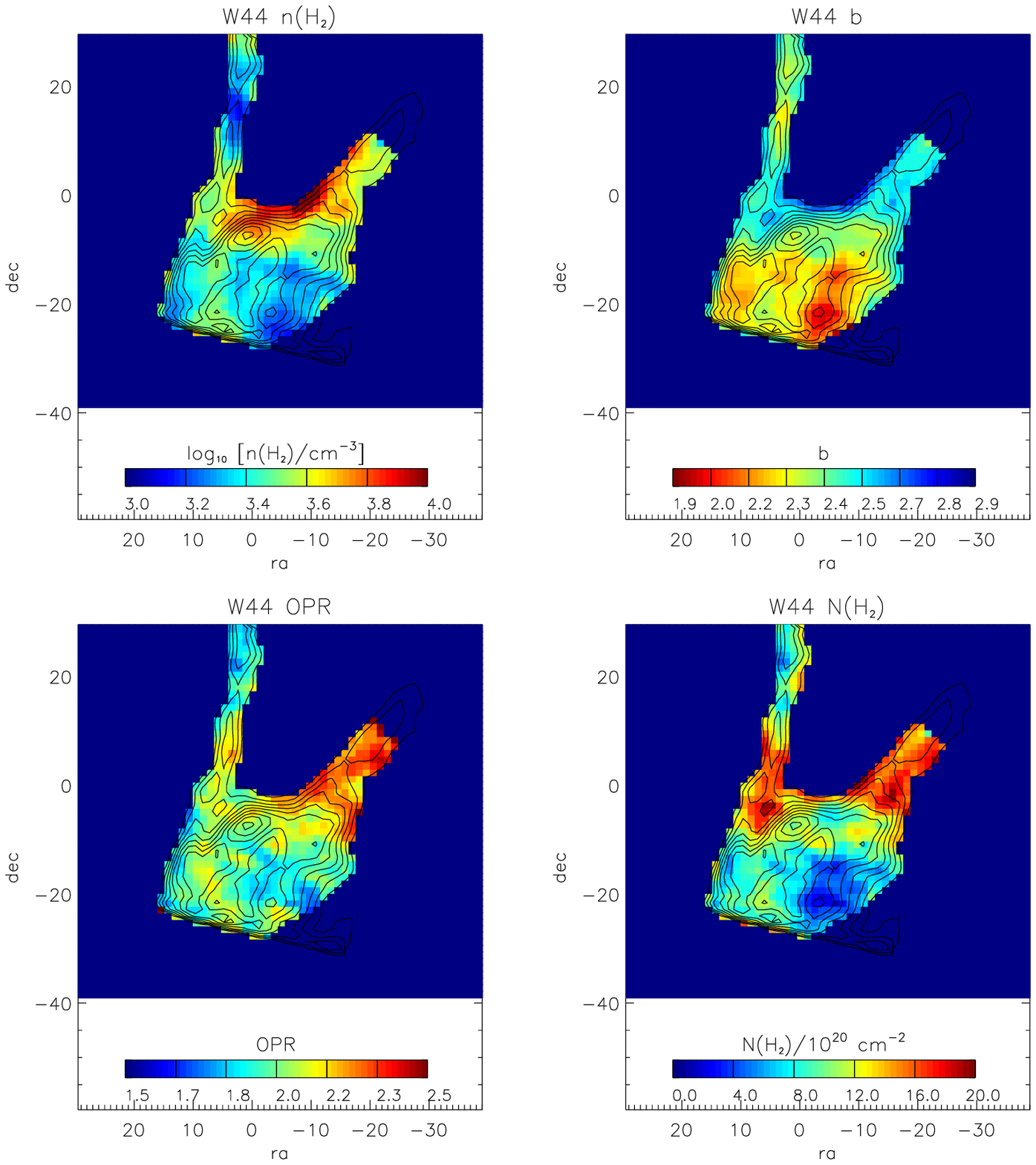}
\caption{Same as Figure \ref{fig:paraW28} but for W44. Regions are
selected with S(5) intensity larger than 8 $\times$10$^{-7}~$W
m$^{-2}$sr$^{-1}$. }\label{fig:paraW44}
\end{figure}

\begin{figure}
\includegraphics[scale=0.7]{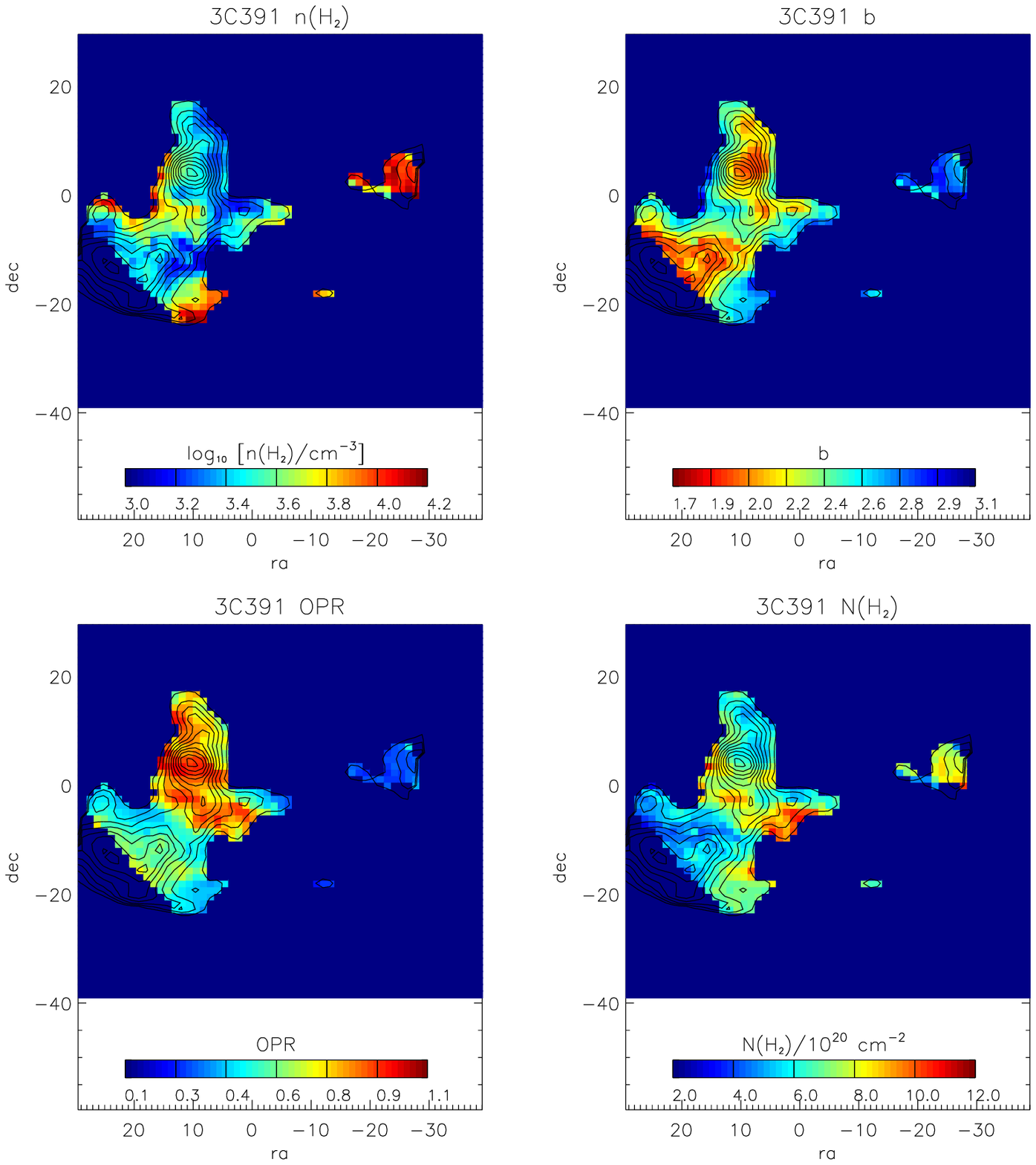}
\caption{Similar as Figure \ref{fig:paraW28} but for 3C391. Regions
are selected with S(5) intensity larger than 7.9 $\times$10$^{-7}~$W
m$^{-2}$sr$^{-1}$. }\label{fig:para3C391}
\end{figure}

\begin{figure}
\includegraphics[scale=0.7]{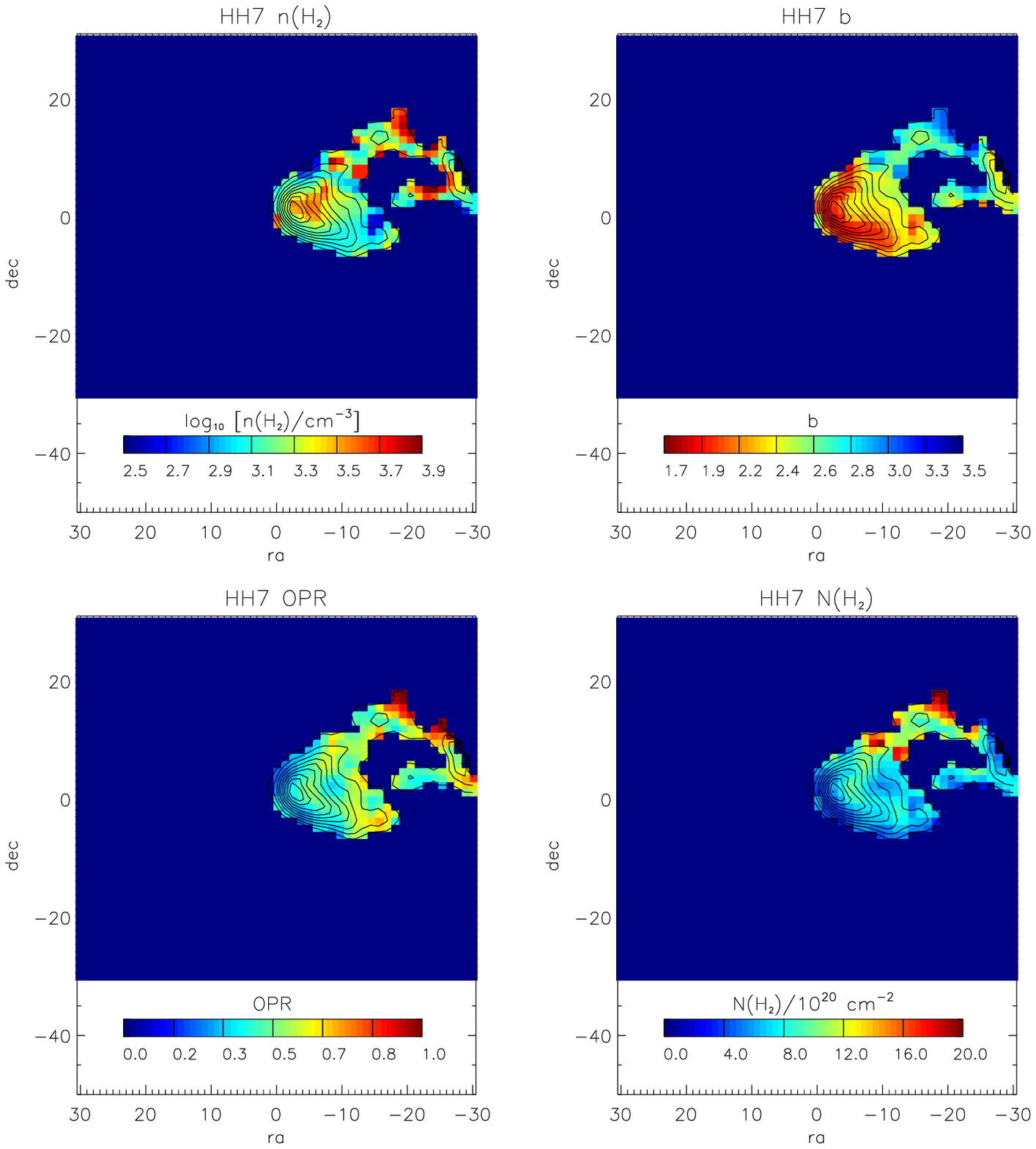}
\caption{Same as Figure \ref{fig:paraW28} but for HH7. Regions are
selected with S(5) intensity larger than 2.0 $\times$10$^{-7}~$W
m$^{-2}$sr$^{-1}$. }\label{fig:paraHH7}
\end{figure}

\begin{figure}
\includegraphics[scale=0.7]{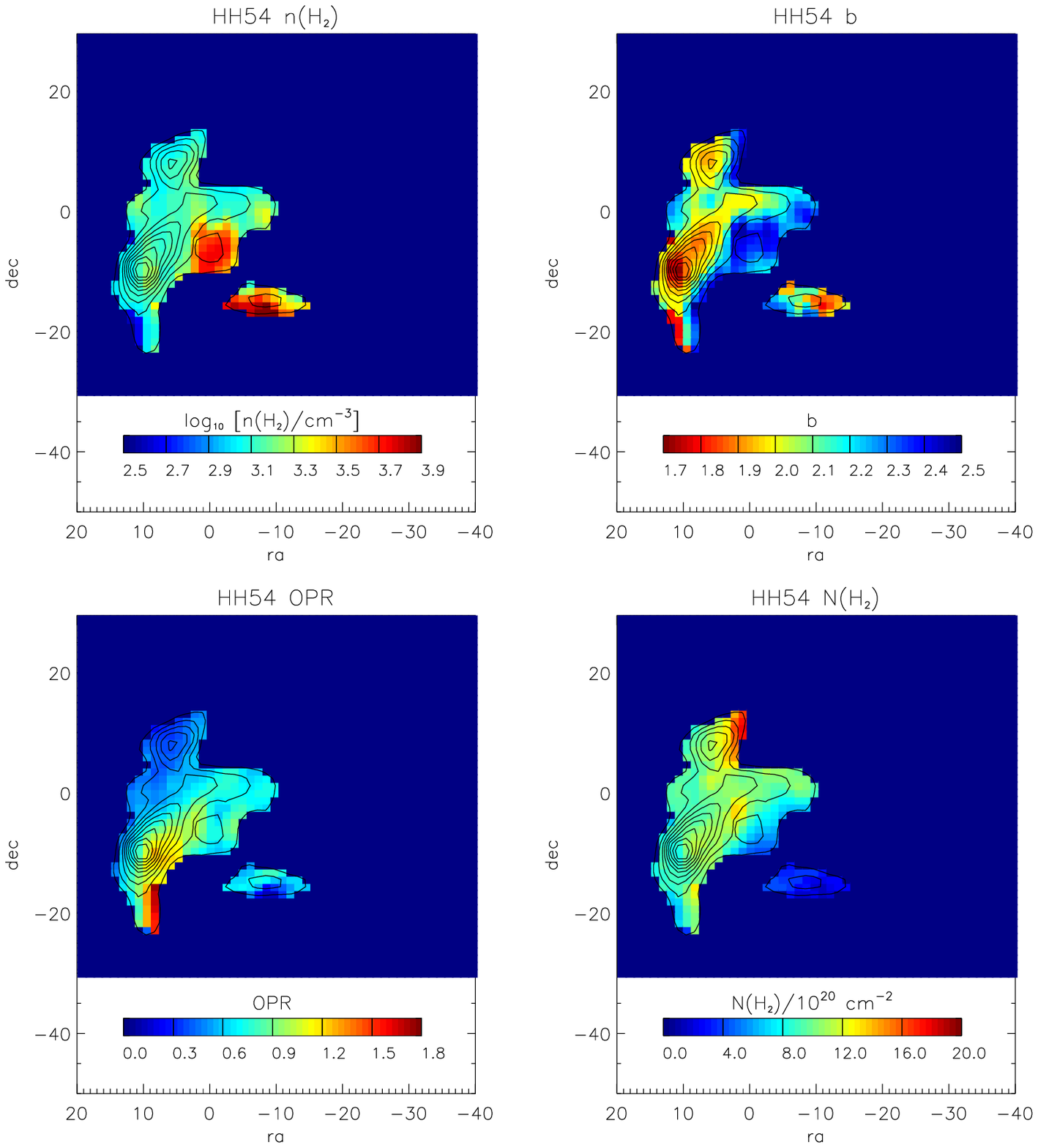}
\caption{Similar as Figure \ref{fig:paraW28} but for HH54. Regions
are selected with S(5) intensity larger than 2.8$\times$10$^{-7}~$W
m$^{-2}$sr$^{-1}$. }\label{fig:paraHH54}
\end{figure}

The spatial distributions of the best-fit parameters -- including the
H$_{2}$ density, the power law index, $b$, the column density of
warm hydrogen above 100 K, and the average OPR -- are shown in Figure
\ref{fig:paraIC443} -- \ref{fig:paraHH54}. The contours of the
brightest H$_{2}$ line, S(5), are superposed. These parameter maps are
obtained by fits to the H$_{2}$ IRS lines only, which have good
signal-to-noise ratios. They are not intended to show the exact
values of the parameters at each position but rather the spatial
variations of the physical conditions in these regions. Here, we show only
the averaged OPR over the column density of H$_{2}$ at every
position, not OPR$_0$ and $n$(H)$\times$$\tau$ separately because
these two parameters, like another pair of parameters, $n$(H$_{2}$)
and $b$, are degenerate in the parameter space. Increasing either of
them will raise the resultant OPR of the gas. Thus the confidence
intervals are wide for these two parameters, especially for
$n$(H)$\times$$\tau$. The much larger uncertainties in the line
intensities at a single pixel, compared with the errors in the map-averaged
intensities, make the derived OPR$_0$ and $n$(H)$\times$$\tau$ even
more poorly constrained and unreliable. That is why we show the
averaged OPR maps instead.

\section{Discussion}

\subsection{The best-fit density}

From Table 1 and Figures 7 -- 12 we see that the H$_2$ pure
rotational emissions detected by IRS of all six sources are
consistent with excitation conditions in gas with $n$(H$_2$) $\sim$
2 -- 4 $\times 10^{3}$ cm$^{-3}$ and temperature index $b\sim$ 2.3
-- 3.1. In the case of W28, W44 and 3C391, the intensities of H$_2$
S(9),  however, suggest a denser region with $n$(H$_2$) 2 -- 2.5
times larger, and including the IRAC band 2 (4.5 $\mu$m) brightness
-- contributed mainly by H$_2$ S(9) to S(12) -- as well as the CO
highly-excited rotational line fluxes, which are considered only in
the case of IC443C and HH54, yields an even higher best-fit density
with $n$(H$_2$) $\sim$ 1 -- 4 $\times10^{4}$ cm$^{-3}$. The latter
density range is closer to estimates from previous studies of
emissions from various species. Snell et al.\ (2005) found that a
preshock density of 3$\times10^{4}$ cm$^{-3}$ for either a slow
J-type or C-type shock in IC443 clump C can account for the observed
H$_2$O, CO, OH and H$_2$ 2 $\mu$m line intensities. For the other
three SNRs -- W28, W44 and 3C391 -- the IRS regions coincide with
the locations of the brightest 1720 MHz OH masers, the presence of
which implies the existence of clumps of OH gas at moderate
temperature 50 -- 125 K and densities $n$(H$_2$) $\sim10^{5}$
cm$^{-3}$ in C shocks (Lockett et al.\ 1999; Wardle \& Yusef-Zadeh
2002). For HH54, the multi-species analysis done by Giannini et al.\
(2006) indicated that an 18 km s$^{-1}$ J-type shock with a
continuous precursor and a density $n$(H$_2$) $\sim 10^{4}$
cm$^{-3}$ matches the H$_2$ vibrational and pure rotational lines,
as well as the CO and H$_2$O emissions observed with {\it ISO}.
Molinari et al.\ (2000) studied the HH 7--11 outflow emissions using
${\it ISO}$ and interpreted the H$_2$, CO and H$_2$O line emissions
as emerging from a mixture of J- and C-type shocks propagating in
gas of density $n$(H$_2$) $\sim 10^{4}$ cm$^{-3}$.

The inconsistency among the best-fit densities estimated from
different molecular species, obtained in the calculations described
above, can be explained by the density fluctuations within the
observed regions.  The clouds may be composed of both moderate
density gas with $n$(H$_2$) $\sim 10^{3}$ cm$^{-3}$ and dense cores
with $n$(H$_2$) $\sim10^{5}-10^{6}$ cm$^{-3}$. Indeed, the density
maps derived from the IRS H$_2$ fluxes exhibit large variations
within the small areas ($\sim~1^{'}\times~1^{'}$) mapped, as shown
in Figures 14 -- 19. The higher critical densities for the H$_2$
S(9) to S(12) transitions, the CO high-$J$ transitions as well as
the H$_2$O lines, which we do not utilize here, make the line
intensities more sensitive functions of density than those of the
H$_2$ IRS transitions. Thus, denser regions contribute more to the
total emission for these transitions of high critical density,
leading to larger density estimates. However, another possibility
also exists that the low-lying and high-lying lines may actually
trace different components of the shock.  Reach et al.\ (2005)
proposed that H$_2$ S(9) can arise largely from the dissociative
part of shock, while H$_2$ S(3) is attributed almost entirely to the
non-dissociative shock. In reality, these two situations may both
exist when part of the highly-excited CO and H$_2$ rotational lines
come from warmer regions where the shock is partially dissociative
and atomic hydrogen becomes an important collisional partner, an
effect neglected in our model.

\subsection{The best-fit temperature distribution index $b$}

The best-fit power-law index $b$, which represents the gas
temperature distribution along the line-of-sight, is in the range
2.3 -- 3.1 according to our fits to the IRS H$_2$ emissions. If H$_2$ S(9) is
also considered, $b$ is enhanced by $\sim$ 0.2, and a further increase of
0.3 -- 0.6 is needed if the IRAC 4.5 $\mu$m band flux or CO lines
are included. The above effects are probably caused by the
degeneracy of the two parameters | $n$(H$_2$) and $b$ | as mentioned
in Section 4; an increase in the best-fit density can be compensated for by
a larger $b$ (i.e. by assuming the presence of less gas at high temperatures).
The best-fit $b$ index
for all six sources is smaller than predictions from a classical bow
shock whose shape can be approximated as parabolic. According to
Smith \& Brand (1990), the effective shock surface area d$A$ with a
perpendicular shock velocity $V_{s}$ is proportional to $V_{s}^{-4}
dV_{s}$, which leads to a power-law index $b$ $\sim 3.8$ if the
relationship between the column density and shock velocity given by
equation B6 in N06 is adopted: $N$(H$_2)$ $\varpropto V_{s}^{-0.75}
$ (NY08). A $b$ index smaller than 3.8 can be caused by a d$A$ which
drops less steeply with velocity $V_{s}$ than does a
parabolic shock.  This would require that the curvature of the
shock front be smaller than
that of a parabola, or more probably, that there exists an admixture of
shocks with different geometries whose shapes vary from planar to bow.

\subsection{ The covering factors within the IRS regions}

The average
column density of the shocked H$_2$ at $T>$ 100 K within the
rectangular areas marked in Figures 1 -- 6 varies from 2$\times$
$10^{20}~$cm$^{-2}$ to 4 $\times$ $10^{21}~$cm$^{-2}$, with the two
Herbig-Haro objects the weakest sources of the total H$_2$
emissions. The length scale defined by $N$(H$_2)$/$n$(H$_2)$, which
should be equal to the product of shock thickness and covering
factor within the regions, is in the range of $10^{16}$ cm --
$10^{18}$ cm.  All sources except HH54 show $N$(H$_2)$/$n$(H$_2)$
above $10^{17}~$cm. The thickness of the shocks in these regions, obtained
from expressions (B6) and (B7) from N06, should be less than
$10^{17}~$cm. The analysis above implies that the covering factor for all
six sources except HH54 is larger than unity; for W28 it is even
as large as $\sim 6$. The high covering factors are not surprising because
we are probably not observing these shocks face-on. The filamentary
structures appearing in part of the maps of W28 and W44 imply the
existence of individual shock fronts seen close to edge-on, as
suggested by Reach et al.\ (2005). Actually, for most of the sources
except HH7, the complicated morphology of the maps suggests
a combination of shocks with different geometry and seen from
different angles. For HH7, the well-defined bow-shaped structure
probably represents a simplified situation. Smith et al.\ (2003)
proposed that a bow shock moving at an angle of $\sim 30 ^{\circ}$
to the line-of-sight is consistent with the H$_2$ line profiles observed at
different positions in HH7.

\subsection{The OPR and the environmental difference between SNRs and Herbig-Haro objects}

Though the physical sizes of the shock structures mapped by IRS for
HH7 and HH54 are $\lesssim 0.1$ times smaller than those for SNRs,
the best-fit density, index $b$ and H$_2$ column density do not
significantly differ between these two classes of source. We noted,
however, that the OPR for hot gas ($T>$ 1000 K) in HH7 and HH54 is
lower than the LTE value $\sim 3$, while for all four SNRs the
departure of the OPR from equilibrium is negligible at that high
temperature. This phenomenon is also reflected in the H$_2$
rotational diagrams, where the zigzag pattern is more apparent for
the two Herbig-Haro objects and is apparent even for the highest
rotational levels. In our model, the equilibrium OPR of hot gas in
the four SNRs requires a best-fit $n$(H)$\times$$\tau$ that is 0.6
-- 3 orders of magnitude larger than that inferred for HH7 and HH54.
This difference may be caused by a different atomic hydrogen density
$n$(H) within gas around SNRs and Herbig-Haro objects. Fast SNR
blast waves driving interstellar shocks with $V_s \geq 100 $~km
s$^{-1}$ can produce strong ultraviolet emissions that are
responsible for the dissociation of H$_2$ in surrounding regions,
while for Herbig-Haro objects the typical shock speeds are observed
to be smaller (Herbig \& Jones 1981;Cohen \& Fuller 1985). Many high
excitation fine structure lines which have been detected previously
toward many SNRs are faint or absent toward Herbig-Haro objects. In
addition to the UV field produced by nearby fast shocks, the X-ray
emission from an SNR interior will also induce dissociation of
pre-shock gas. These all imply that the molecular gas associated
with Herbig-Haro objects probably subject to weaker
photodissociation, which results in less H and a lower efficiency of
para-to-ortho conversion.  We note, however, that for measurements
of near-infrared H$_2$ ro-vibrational transitions toward Herbig-Haro
objects, the OPR values obtained are in most cases consistent with 3
(Smith, Davis \& Lioure 1997). With energy levels lying above 6000
K, these vibrationally-excited lines originate mostly from the
hottest part of the gas, which probably has a higher atomic
fraction. These two factors -- high temperature and high atomic H
fraction -- add up to a fast, efficient ortho-para conversion.

\subsection{Maps of the parameters}

The maps of best-fit parameters in Figures 14 -- 19 were derived from
the IRS H$_2$ fluxes only.  So they may not represent the real
average value at each pixel, but we expect that they carry useful
information about spatial variations in the physical conditions in
these regions. Maps of all six sources exhibit a spatial variation
in $n$(H$_2$) larger than a factor 5,  and a variation in $b$ larger than
1.  The H$_2$ densities derived for a single pixel vary from
$6\times 10^{2}$ cm$^{-2}$ to $10^{4}$ cm$^{-3}$, and the index $b$
ranges from 1.6 to 3.3.

After comparing the IRS H$_2$ spectroscopic maps with the parameter
maps, we found that the maps of the temperature distribution index
$b$ look most similar to the distributions of mid-lying H$_2$
emissions including S(3), S(4) and S(5). To show this similarity, we
superpose the H$_2$  S(5) emission contours on these images. This
fact implies that the emission in these mid-lying transitions is
more strongly dependent on $b$ than on the density.  In other words,
these transitions, H$_2$ S(3) -- S(5), trace mainly the hottest
components of the gas. Since the gas temperature distribution at
each position is determined largely by the shock velocity, these
H$_2$ emissions may also serve as a good tracer of the local
effective shock velocity. In HH7 for example, the S(3) -- S(5) line
intensities appear strongest near the head of the bow, where $V_s$
reaches its maximum.  On the other hand, although the S(6) and S(7)
emissions are also strongly affected by the gas temperature, they
show more dependence upon $n$(H$_2$) compared to other lower-lying
transitions. The critical densities for the excitation of S(6) and
S(7) are higher than $10^{5}$ cm$^{-3}$ at the typical temperatures
of relevance here. Thus, regions of enhanced density show up as
clumpy features within the S(7) map for HH7. Finally, we note that
the derived column density
 of H$_2$ at $T>$ 100 K is mostly determined by the intensities of the low-lying transitions,
 especially S(0). The S(0) emission arises mainly from lower temperature gas with $T < $ 500 K, which contributes most to the total $N$(H$_2$) for the power-law temperature distribution that we assume.

 \section{Summary}

1. We have studied the physical conditions within shock-excited
molecular gas associated with  IC443C, W28, W44, 3C391, HH7 and
HH54.  We mainly used the H$_{2}$ S(0) to S(7) spectral line maps
obtained by IRS on $Spitzer$ to constrain the best-fit parameters.
IRS observations of HD emissions, IRAC band 2 (4.5 $\mu$m) intensity maps, and
{\it ISO} measurements of the
H$_{2}$ S(9)/S(3) ratio and the CO high-lying rotational lines (from $J$ =
14 -- 13 to $J$ = 20 -- 19) are also used when available to provide
additional constraints.

2. A comparison between the IRS H$_2$ emission distribution and the IRAC
maps for IC443C shows the IRAC band 2, 3 and 4 intensities are attributable
almost entirely to H$_{2}$ pure rotational emissions. IRAC band
2 gives us access to the high-lying H$_{2}$ transitions S(9) to S(12) which
are not available from IRS observations. For HH54, the similarity
between the IRS H$_2$ and IRAC maps implies these IRAC band fluxes may
come mostly from H$_2$ emissions as well. We assumed that the HH54 IRAC
band 2 intensity is dominated by H$_2$ emissions and used it as an extra
diagnostic in the model.  For the other four sources the IRAC maps
show either a strong continuum component from PAHs or dust or are
heavily polluted by point sources.

3. We adopted a power-law temperature distribution for the shocked
gas, with the column density of gas at temperature between $T$ and
$T$ + $dT$ assumed to be proportional to $T^{-b}$, where $T$ ranges
from 100 to 5000 K. The molecular line intensities are then modeled
under the assumption of statistical equilibrium. We have checked the
validity of this approximation for transitions of the three species
H$_{2}$, HD and CO. Our calculations show the departure from
statistical equilibrium for those rotational states involved in our
calculation is negligible under all plausible density conditions in
molecular shocks.

4. The best-fit densities determined from the IRS H$_{2}$ pure rotational
lines S(0) to S(7) for all six sources are consistent with the
excitation conditions in gas with $n$(H$_2$) $\sim$ 2 -- 4
$\times 10^{3}$ cm$^{-3}$. The intensities of H$_2$ S(9), however,
require an environment 2 -- 2.5 times denser. In the case of IC443C
and HH54, where the highly-excited CO rotational line
intensities measured by {\it ISO} are reliable and IRAC 4.5 $\mu$m band fluxes were
considered, we found the gas density determined by
 including all the data above is even larger: $n$(H$_2$)
$\sim$ 1 -- 4 $\times10^{4}$ cm$^{-3}$. This inconsistency can
be explained by density fluctuations within the observed regions.
However, it is also possible that the low-lying
and high-lying lines originate from different components of the
shock.

5. For all six sources the best-fit power-law index $b$ derived
from IRS H$_{2}$ S(0) to S(7) is in the range of 2.3 -- 3.1. If H$_2$
S(9) is also considered, $b$ is enhanced by $\sim$ 0.2, and a further
increase of 0.3 -- 0.6 is needed if the CO lines or IRAC 4.5 $\mu$m
band fluxes are also included. The  best-fit $b$ index is smaller
than predictions from a classical parabolic bow shock, which leads to
a power-law index $\sim 3.8$. It can be understood if the average
curvature of the shock front is smaller than that of a parabola |
or if there exists an admixture of shocks whose shapes vary from planar
to bow.

6. The OPR for hot gas ($T> $ 1000 K) in all four SNRs is fairly
close to the LTE value of $3$, while for the two Herbig-Haro objects it is confirmed
to be less than the LTE value even at $T>$ 1000 K. This difference
may be caused by different preshock atomic hydrogen densities $n$(H)
within gas around SNRs and Herbig-Haro objects. SNRs may be
subject to heavier UV photodissociation and therefore produce more atomic
hydrogen in the gas, which is the dominant collisional partner in
the para-to-ortho conversion process in molecular shocks.

7. Unlike the OPR, the best-fit density, power-law index, $b$, and H$_2$
column density do not differ significantly
between SNRs and Herbig-Haro objects.  It should be noted that
the acceptable ranges of the fitted parameters are actually large
because they are degenerate in the parameter space ($n$(H$_2$) versus
$b$ and OPR$_0$ versus $n$(H)$\times\tau$).

8. Given the observed IRS H$_2$ fluxes, we obtain upper limit on the CO
abundance | $n$(CO)/$n$(H$_2$) | within the 80$''$ {\it ISO}/LWS beam of
$1.5\times 10^{-3}$ and $ 5.5\times 10^{-4}$ for IC443C and HH54,
respectively.  Assuming that the CO emission distribution is similar
to that of the IRAC bands (or the H$_2$ emission), we derive a rough estimate
for the CO abundance of  $n$(CO)/$n$(H$_2$) $\sim$ 3 --
5$\times 10^{-5}$ for IC443C and $n$(CO)/$n$(H$_2$) $\sim$ 2 --
4$\times 10^{-5}$ for HH54.

9. Parameter maps derived from the H$_2$ S(0) to S(7) lines for all six sources
exhibit a spatial variation in $n$(H$_2$) larger than a factor 5 and a variation in $b$ larger than 1.
 The density, $n$(H$_2$), varies from $6\times10^{2}$ to $10^{4}$ cm$^{-3}$, and the index $b$ ranges from 1.6 to 3.3.

10.  Our maps of the best-fit parameters
 indicate that the mid-lying H$_2$ emissions | S(3) to S(5) | trace the hot
 component of the gas. On the other hand, the excitation of high-lying
 transitions, including S(6) and S(7), is more sensitive to the density of the gas.
The spatial distribution of the H$_2$ column density with $T>$ 100 K
is determined mainly by the lowest-lying transitions, particularly S(0).

\appendix
\section{Evolution of level populations of H$_{2}$, HD,
CO from non-equilibrium state}

We present here a simple analysis of the relaxation timescale for
the three species in shocks by solving the time-dependent population
transfer equations
\begin{equation}
\sum_{J'}\left (C_{J'\rightarrow J}+A_{J'\rightarrow J}\right
)f_{J'}-f_J \cdot \sum_{J'}\left (C_{J\rightarrow
J'}+A_{J\rightarrow J'}\right )={df_J \over dt},
\end{equation}
similar to equation (1) but with a time-dependent term.

\begin{figure}
\includegraphics[scale=0.7]{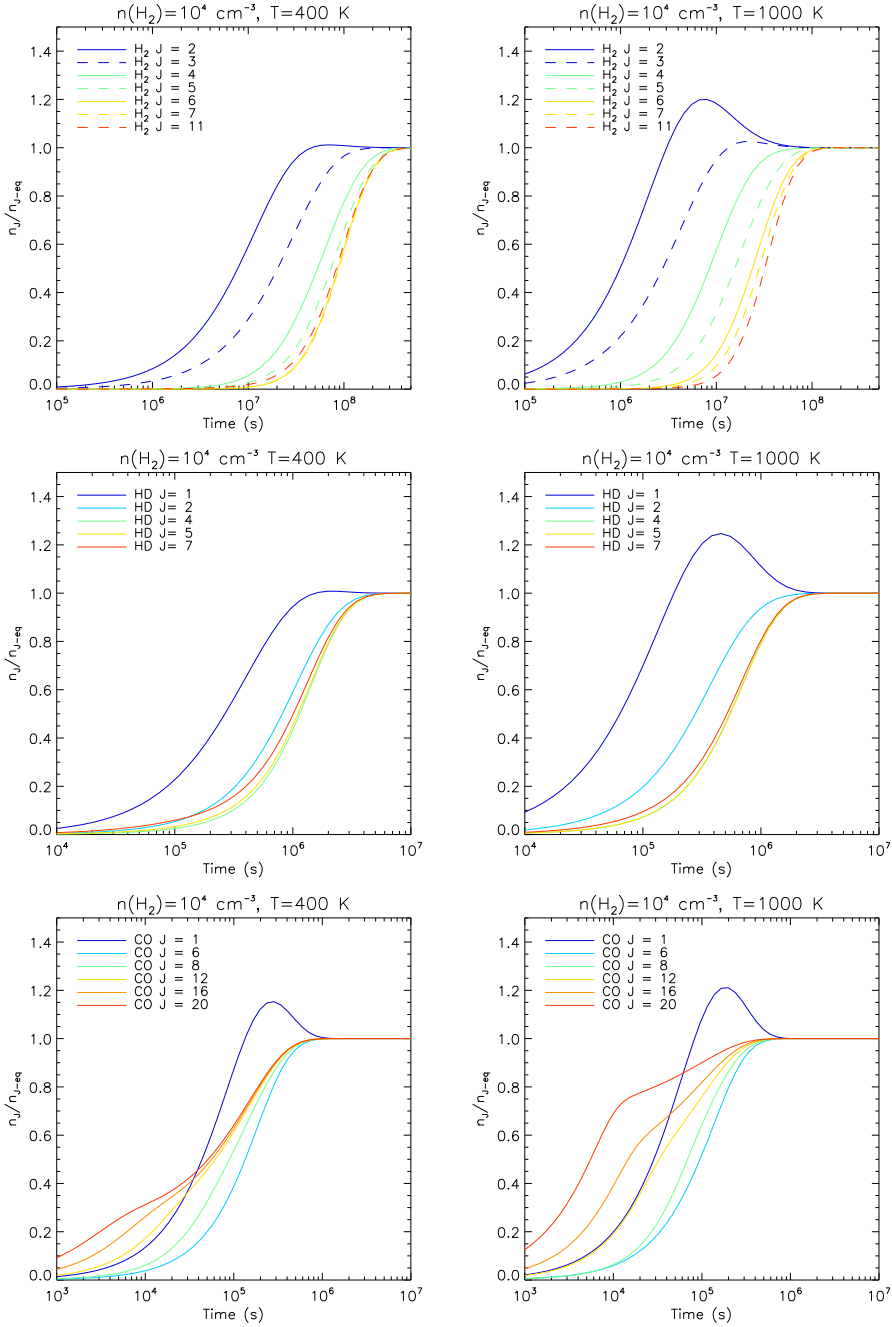}
\caption{The evolution of level populations of H$_2$, HD and CO at
density $n$(H$_2$) = 10$^{4}$ cm$^{-3}$. $n_J$ denotes the level
population density for level $J$ and $n_{J-eq}$ is the value at
statistical equilibrium. The left panels are for $T$ = 400 K and
right panels for $T$ = 1000 K, values which are consistent with
typical temperatures of the warm and hot components (see
text).}\label{fig:Allfevon4}
\end{figure}

To solve the equations, all the molecules are assumed to be
initially in the lowest quantum state. Here we treat ortho- and
para- H$_{2}$ as distinct species due to the remarkably low
efficiency of the para-to-ortho conversion processes, especially
when compared with that of collisional and radiative transitions.
Thus all para-H$_{2}$, HD and CO are put at $J = 0$ and all
ortho-H$_{2}$ are at $J = 1$ at the beginning. The time evolution of
the level populations for the three molecules at constant density
$n$($H_{2}$) = 10$^{4}$ cm$^{-3}$ and constant temperature $T$ = 400
K or $T$ = 1000 K (typical temperatures for the warm and hot
components fitted by N06 \& N07) are presented in Figure
{\ref{fig:Allfevon4}}. The collisional partners were assumed to be
molecular hydrogen and helium only. An average velocity
gradient of 2 $\times$ 10$^{-10}$ cm s$^{-1}$/cm and a
$n$(CO)/$n$(H$_2$) ratio of 5$\times10^{-5}$ were adopted in calculating
the CO optical depth. We note however, that these values do not
affect greatly the evolution timescale.

From equation (A1), it is quite straightforward to see that the
evolution depends on two processes | radiative decay and collisional
excitation (and de-excitation). We can define a characteristic time
$t_{1/2}$, at which the population of a certain level $J$ reaches
one-half of that in statistical equilibrium. Apparently, at the low
density limit where $n$(H$_{2}$) is smaller than the critical
density for the transition between the lowest two levels, radiative
processes dominate. Here, $t_{1/2}$ will approach 1/A$_{J\rightarrow
J-1}$. For the four species para-H$_{2}$, ortho-H$_{2}$, HD and CO,
the critical densities at 400 K are of the order of 10 cm$^{-3}$,
10$^{2}$ cm$^{-3}$, 10$^{2}$ cm$^{-3}$ and 10$^{3}$ cm$^{-3}$. By
contrast, for extremely dense gas which is allowed to reach local
thermal equilibrium (LTE), $t_{1/2}$ is determined by the inverse of
the collisional excitation rate $\sim$ 1/C$_{J-1\rightarrow J}$,
where C$_{J-1\rightarrow J}$ is proportional to $n$(H$_{2}$). In
most cases with density between the low and high limits, $t_{1/2}$
lies between 1/A$_{J\rightarrow J-1}$ and 1/C$_{J-1\rightarrow J}$.
More generally, the time required to reach equilibrium, however, is
determined by the rotational state that reaches equilibrium most
slowly. We define a relaxation time as the time required for all
level populations to achieve values within 5$\%$ of those attained
in statistical equilbrium.  For this definition, we include the
first 18 levels of H$_2$, the first 8 levels of HD and the first 21
levels of CO. From figure \ref{fig:Allfevon4} we see that at
$n$(H$_{2}$) = 10$^{4}$ cm$^{-3}$ and $T$ = 400 K, H$_{2}$, HD and
CO have relaxation times of 2.1$\times$10$^{8}$ s, 3$\times$10$^{6}$
s and 4.6$\times$10$^{5}$ s respectively. For a hotter gas component
with $T = 1000$~K, the relaxation times are reduced to
8$\times$10$^{7}$ s, 1.6$\times$10$^{6}$ s and 3.3$\times$10$^{5}$
s. Assuming a shock velocity in the range of 10 -- 20 km s$^{-1}$,
and a typical H$_2$ column density of $\sim$ 3$\times$10$^{20}$
cm$^{-2}$ given by equation (B7) in N06 for planar shocks, the fluid
will spend a flow time larger than 1.5$\times$ 10$^{10}$ s passing
through the whole shock affected region, much longer than the
relaxation timescale defined above.

\begin{figure}
\includegraphics[scale=0.7]{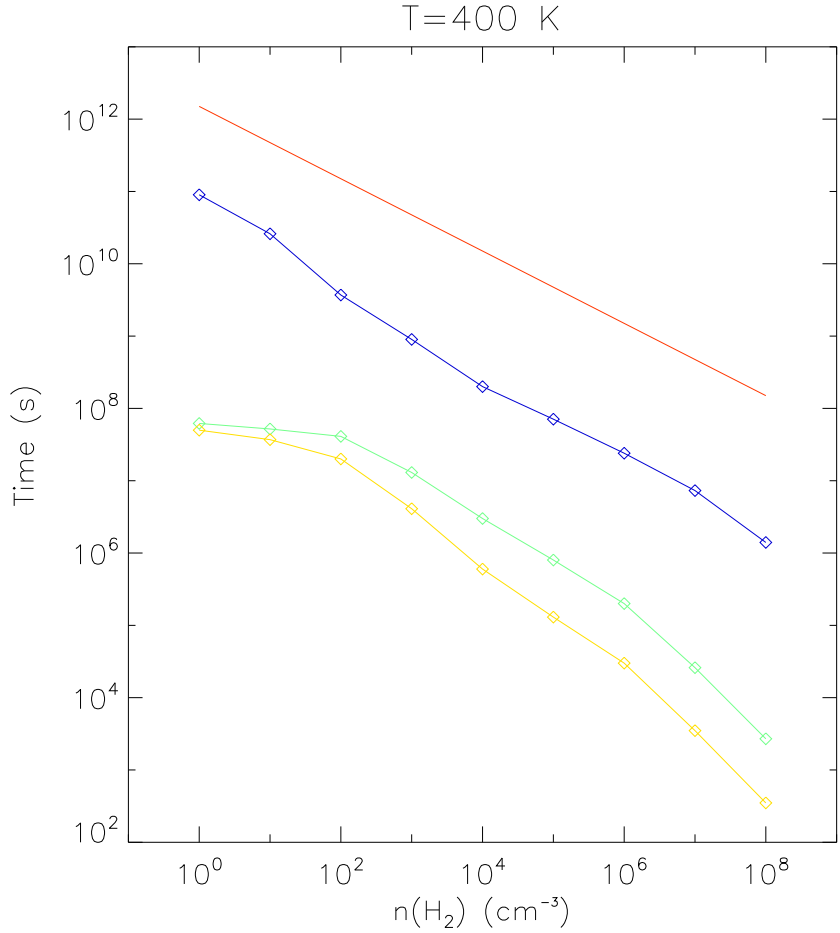}
\caption{The dependence of the relaxation times upon H$_2$ density
at 400 K. We define the relaxation time as the time needed for the
level populations to attain values within 5$\%$ of the those
obtained in statistical equilibrium. For H$_{2}$, we take the larger
of the timescales calculated separately for para- and ortho-H$_2$.
The blue, green and yellow squares represent values for H$_{2}$, HD
and CO respectively. The red line marks the flow time; here we
assumed $t_{flow}$ to be proportional to $n$(H$_2$)$^{-0.5}$ and
equal to $\sim$ 1.5$\times$ 10$^{10}$ s at $n$(H$_2$)= 10$^{4}$
cm$^{-3}$.}\label{fig:alldensities}
\end{figure}

If we assume a shocked H$_2$ column density proportional to
$n$(H$_2$)$^{0.5}$ as given by equation (B7) in N06, the flow time
$t_{flow}$ $\sim$ $N$(H$_2$)/($n$(H$_2$)$V_S$) will be proportional
to $n$(H$_2$)$^{-0.5}$. Given a t$_{flow}$ $\sim$ 1.5$\times$
10$^{10}$ s at $n$(H$_2$)= 10$^{4}$ cm$^{-3}$, a comparison between
relaxation times for H$_{2}$, HD, CO and the flow time at various
densities with a typical shock velocity $\sim$ 20 km s$^{-1}$ is
given in figure \ref{fig:alldensities}. It is apparent that, for
modeling the molecular transitions accessible to infrared
observatories including IRS, IRAC and ISO, the departure from
statistical equilibrium for those levels involved is always
negligible under all possible density conditions in molecular
shocks.

\clearpage

\begin{deluxetable}{lccccccc}
\tabletypesize{\scriptsize}\tablecolumns{7} \small\tablewidth{0pt}
\tablecaption{Observed fluxes and average line intensities}
\tablehead{ \colhead{Species}
&\colhead{IC443C}&\colhead{W28}&\colhead{
W44}&\colhead{3C391}&\colhead{ HH7}&\colhead{HH54}}
\startdata H$_2$ S(0) 28.22 $\mu m$ &0.14\tablenotemark{a}  &0.37  &0.079 &0.13  &0.093 &0.058\\
H$_2$ S(1) 17.04 $\mu m$ &3.45   &2.55 &1.16  &0.69  &0.39 &0.40\\
H$_2$ S(2) 12.28 $\mu m$ &4.42  &3.45 &1.39 &0.96  &1.65 &1.79 \\
H$_2$ S(3) 9.67 $\mu m$ &19.95   &6.35  &3.44 &1.88  &2.04 &2.75\\
H$_2$ S(4) 8.03 $\mu m$ &8.05   &2.06  &2.47 &1.60  &2.16 &2.95 \\
H$_2$ S(5) 6.91 $\mu m$ &23.57   &6.34 &8.46 &8.88 &2.71 &3.61\\
H$_2$ S(6)\tablenotemark{b} 6.10 $\mu m$ &5.08   &... &... &... &1.10 &1.80\\
H$_2$ S(7) 5.51 $\mu m$ &11.38   &2.22 &4.51 &3.51  &1.48 &1.89\\
\tableline
HD R(3) 28.50 $\mu m$ &0.040  &... &... &... &0.012 &0.0093\\
HD R(4) 23.03 $\mu m$ &0.017  &...  &... &...  &0.0071 &0.0075\\
\tableline
H$_2$ S(9)/S(3)\tablenotemark{c} &...    &0.12 &0.53 &0.67 &...  &...\\

IRAC Band2(4.5 $\mu m$) &6.91\tablenotemark{d}  &2.11 &4.54 &10.41  &2.58 &1.85\\
\tableline
CO $J$=14-13 186.00 $\mu m$ &... &... &... &9.0$\pm$2.2\tablenotemark{f} &6.6$\pm$2.1  &6$\pm$1\\
CO $J$=15-14 173.63 $\mu m$ &12.9$\pm$4.9\tablenotemark{e}   &$<15$\tablenotemark{f}   &...     &28.8$\pm$2.3   &8.5$\pm$2.8 &6$\pm$1 \\
CO $J$=16-15 162.81 $\mu m$ &11.7$\pm$2.2   &$<5$    &$<5$\tablenotemark{f}   &10.0$\pm$1.7  &5.7$\pm$1.5 &3.9$\pm$0.2 \\
CO $J$=17-16 153.27 $\mu m$ &6.7$\pm$0.9  &... &... &10.3$\pm$2.8  &8.4$\pm$3.0 &2.2$\pm$0.3 \\
CO $J$=18-17 144.78 $\mu m$ &...     &...  &... &$<17$  &...  &2.1$\pm$0.3\\
CO $J$=19-18 137.20 $\mu m$ &...    &...  &... &$<17$         &...  &$<1$\\
CO $J$=20-19 130.37 $\mu m$ &...     & ...  &... &$<17$ &...  &0.9$\pm$0.3\\

\enddata
\tablenotetext{a}{The unit of IRS H$_2$ and HD line intensities is
10$^{-7}$W m$^{-2}$sr$^{-1}$.} \tablenotetext{b}{For W28, W44 and
3C391 the IRS H$_2$ S(6) is blended with strong 6.2 $\mu$m PAH
feature and can not be measured.} \tablenotetext{c}{The S(9)/S(3)
ratios for W28, W44 and 3C391 are derived from the H$_2$ line fluxes
measured by SWS observations (Reach \& Rho 1998).}
\tablenotetext{d}{The unit of IRAC band intensity is MJy sr$^{-1}$
(10$^{-20}$W m$^{-2}$sr$^{-1}$Hz$^{-1}$).  } \tablenotetext{e}{The
unit of CO line fluxes for IC443C, HH54, HH7 is {10$^{-16}$W
m$^{-2}$}.} \tablenotetext{f}{The unit of CO line intensities for
W28, W44 and 3C391 is {10$^{-9}$W m$^{-2}$sr$^{-1}$}.}

\end{deluxetable}

\begin{deluxetable}{lccccccc}
\tabletypesize{\scriptsize}\tablecolumns{7} \small\tablewidth{0pt}
\tablecaption{Observed fluxes and average line intensities}
\tablehead{ \colhead{Best-fit parameters}
&\colhead{IC443C}&\colhead{ W28}&\colhead{ W44}&\colhead{
3C391}&\colhead{ HH7}&\colhead{HH54}} \startdata

\tableline Fits to IRS H$_2$ lines only\\

$log_{10}$[$n$(H$_2$)/cm$^{-3}$]       &3.36 (3.1 -- 3.9)\tablenotemark{a}  &3.42 (3.2 -- 4.1) &3.63 (3.2 -- 4.5) &3.62 (3.2 -- 4.5)  &3.39 (2.9 -- 4.2) &3.30 (3.0 -- 3.7)\\
Power law index, b                &2.29 (1.9 -- 2.7)    &3.07 (2.7 -- 3.7)   &2.47 (2 -- 3)   &2.48 (2 -- 3.1)  &2.54 (2.2 -- 3.1)  &2.17 (1.8 -- 2.6)  \\
OPR$_0$                              &3\tablenotemark{b}  &1.23 ($\leq$2.7)    &1.98 ($\leq$3) &0.255($\leq$1.8)  &0.517 (0.2 -- 1.1)  &0.65 ($\leq$1.2) \\
$log_{10}$[$n$(H)$\times\tau/cm^{-3}$yr] &...
&4.43 (...)\tablenotemark{c} &3.88 (...) &6.38 ($\geq$2.3)  &3.25 (2.6 -- 4.2) &3.01 ($\leq$3.7)\\
$log_{10}$[$N$(H$_2$)/cm$^{-2}$]       &21.13  &21.51  &20.87  &20.83   &20.63 &20.38\\

\tableline Fits to all reliable features\\
log$_{10}$[$n$(H$_2$)/cm$^{-3}$]       &4.1 (3.2 -- 4.7)  &3.83 (3.2 -- 4.4) &3.92 (3.4 -- 4.4) &3.91 (3.5 -- 4.4) &3.41 (2.9 -- 4.2) &4.51 (3.4 -- 5.5)  \\
Power law index, b                  &2.78 (2.2 -- 3.2)  &3.36 (2.8 -- 3.9)   &2.65 (2.2 -- 3.1)   &2.68 (2.2 -- 3.1)  &2.55 (2.2 -- 3.1)   &2.99 (2.4 -- 3.3)  \\
OPR$_0$                              &3  &1.44 ($\leq$3)    &2.23 ($\leq$3) &0.493 ($\leq$2)  &0.520 (0.2 -- 1.1)  &0.787 ($\leq$1.5) \\
log$_{10}$[$n$(H)$\times\tau$/cm$^{-3}$yr] &...
&4.43 (...) &3.48 (...) &6.14 ($\geq$3.3) &3.25 (2.6 -- 4.2) &2.82 ($\leq$3.6)\\
log$_{10}$[$N$(H$_2$)/cm$^{-2}$]       &21.32  &21.61  &20.95  &20.92  &20.63 &20.69\\

\enddata
\tablenotetext{a}{The $95.4\%$ confidence intervals are shown in the
parentheses.} \tablenotetext{b}{The H$_2$ rotational diagram for
IC443C shows no apparent departure from the equilibrium value of
ortho-to-para ratio and is consistent with OPR = 3.}\tablenotetext{c}{The $95.4\%$ confidence
limit is not effective here -- it covers the whole region where $n$(H)$\times\tau$/cm$^{-3}$yr $>0$.}
\end{deluxetable}

\clearpage

\clearpage

\clearpage

\clearpage

\clearpage

\clearpage

\clearpage

\clearpage

\clearpage

\clearpage

\clearpage

\clearpage

\clearpage

\clearpage

\clearpage

\clearpage

\clearpage

\clearpage

\clearpage

\clearpage

\clearpage

\end{document}